\newcommand{\secref}[1]{Sec.~\ref{#1}}

\newcommand{\subsecref}[1]{Subsec.~\ref{#1}}
\newcommand{\figref}[1]{Fig.~\ref{#1}}
\newcommand{\ud}{\mathrm{d}}
\newcommand{\bo}[1]{\boldsymbol{#1}}

\documentclass[aip,jcp, preprint]{revtex4-1}
\usepackage{epsfig}
\usepackage[latin1]{inputenc}
\usepackage{amssymb,amsmath,amsfonts}
\usepackage[pdf]{psfrag}
\newsavebox{\IBox}

\begin{document}

\title{Diffusion of multiple species with excluded-volume effects} 

\author{Maria Bruna}
\email{bruna@maths.ox.ac.uk}
\affiliation{Mathematical Institute, University of Oxford, 24-29 St. Giles', Oxford OX1 3LB, United Kingdom}

\author{S. Jonathan Chapman}
\email{chapman@maths.ox.ac.uk}
\affiliation{Mathematical Institute, University of Oxford, 24-29 St. Giles', Oxford OX1 3LB, United Kingdom}

\date{\today}

\begin{abstract}
Stochastic models of diffusion with excluded-volume effects are used to model many biological and physical systems at a discrete level. The average properties of the population may be described by a continuum model based on partial differential equations. In this paper we consider multiple interacting subpopulations/species and study how the inter-species competition emerges at the population level. Each individual is described as a finite-size hard core interacting particle undergoing Brownian motion. The link between the discrete stochastic equations of motion and the continuum model is considered systematically using the method of matched asymptotic expansions. The system for two species leads to a nonlinear cross-diffusion system for each subpopulation, which captures the enhancement of the effective diffusion rate due to excluded-volume interactions between particles of the same species, and the diminishment due to particles of the other species. This model can explain two alternative notions of the diffusion coefficient that are often confounded, namely collective diffusion and self-diffusion. Simulations of the discrete system show good agreement with the analytic results.
\end{abstract}

\pacs{35C20, 35K55, 35Q84, 60J70, 82C22}

\maketitle 

\section{Introduction}

Stochastic models describing how interacting individuals give rise to collective behavior have become a widely used tool across disciplines---ranging from biology to physics to social sciences.\cite{Helbing:2001ue,Murray:2009ej,vandykeparunak1998agent} Despite their conceptual simplicity, particle-based models can be computationally intractable for large systems of interacting particles, as is often the case in practical applications. In such cases, a continuum population-level description based on partial differential equations that can capture the overall population density becomes attractive. The challenge is then to predict the correct population-level description of the key attributes at the particle level (such as interactions between individuals and evolution rules). 

In particular, the model of diffusive particles with hard-core repulsive (or steric) interactions is relevant to many applications, such as colloidal systems,\cite{JabbariFarouji:2012cr}  ion transport,\cite{Boda:2011gc,Hille:2001tw} diffusion through polymers,\cite{Kruger:2009dv} biological cell populations\cite{Sun:2007dc,Lushnikov:2008jt} and animal behavior.\cite{John:2004gq}
In our previous work \cite{Bruna:2012cg} we considered the diffusion under an external force of $N$ hard spheres in a bounded domain $\Omega \in \mathbb R^d$, $d=2,3$ (of typical nondimensional diameter one). The particles were taken to be identical with nondimensional diameter $\epsilon \ll 1$ and diffusivity $D$. Starting from the particle-level description, we used a method based on matched asymptotic expansions for a small but finite volume fraction to obtain the continuum model. The result is a nonlinear diffusion equation for the one-particle probability density function $p({\bf x}, t)$,
\begin{align}
\label{fpN_one} 
\frac{\partial p}{\partial t}({\bf x}, t) &= \bo \nabla_{{\bf x}} \cdot \left \{ D \bo \nabla_{{\bf x}}\! \left [ p +  (N-1) \alpha \epsilon ^d \, p^2 \right ] -  {\bf  f}({\bf x}) \, p \right \} \qquad \text{in} \qquad \Omega,
\end{align}
where $\alpha = 2(d-1)\pi/d$ and ${\bf f}({\bf x})$ is the nondimensionalized force (drift). Since the nonlinear term is positive for $N>1$, we find that excluded-volume effects enhance the overall collective diffusion rate.

The case of several different types of  particle is relevant in many practical problems but has been paid far less attention by the mathematical community so far.\cite{Burger:2010gb} For example, multiple populations of interacting agents appear in traffic flow with heterogeneous agents,\cite{Helbing:2001ue} pedestrian or animal motion (e.g. ants going in opposite directions \cite{John:2004gq}) and  cellular tumor invasion (cancer and normal cells).\cite{Sander:2002jb} They are also important in ion transport through membrane channels, as in many applications ions can be heterogeneous.\cite{Gillespie:2002tj} Another application is the extreme case in which one of the populations is motionless and blocks the motion of the others; for example, anomalous diffusion in cell membranes due to  obstruction (from, e.g, the membrane skeleton mesh, fixed proteins, or lipid rafts).\cite{NicolauJr:2007dy,Saxton:1994hk}

Much of the effort to describe multiple species with size-exclusion processes has been directed at on-lattice models, in which the motion of particles is restricted to taking place on a lattice and one defines certain hopping rules between lattice sites to account for the particle motion and interactions. A common approach is to take a continuum limit of the discrete model and obtain a  partial differential equation (PDE) describing the average occupancy of the agent population.\cite{Baker:2012dv} For example, this strategy applied to a multi-species motility model based on simple exclusion process with drift (or bias) leads to a system of nonlinear advection--diffusion equations.\cite{Burger:2010gb,Simpson:2009gi} More complicated rules have also been considered; namely, particles that can bind to sites and interact not only with each other but also with a confined channel-like domain,\cite{Henle:2008fv} or \emph{myopic} agents (in which the hopping probability depends on the number of unoccupied nearest-neighbor sites). \cite{Landman:2011ek} Recently, Penington \emph{et. al.} have considered a generalization of these specific on-lattice models to incorporate general interactions, and derived the associated continuum models systematically. \cite{Penington:2011ga} 

Here, we are interested in off-lattice models, where particles each undergo a continuous Brownian motion. We extend our previous work for identical hard-core interacting particles to the case when two types of particles are present. We call these two \emph{species} the blues and reds after Ref.~\onlinecite{Burger:2010gb}. Specifically, we allow for each subpopulation to have a different number of particles with different sizes and diffusivities and to be under a different external force. 

This modeling of such system of interacting particles is typically based on a microscopic approach using $N$-coupled Brownian motions, where $N$ is the total number of particles, or a macroscopic approach using partial differential equations (PDEs). For two species, the latter consists of a system of PDEs for the two subpopulation densities. As mentioned before, microscopic models generally require many computationally intensive simulations to gain understanding of population-level behavior, and can become impractical to use. This is why continuum models are a very useful tool, and there is a lot of interest in predicting the correct macroscopic description of the particle-level attributes. This can be a very challenging task, especially when nontrivial interparticle interactions are present in the system. This is why continuum models are often defined phenomenologically (that is, written directly at the continuum level rather than derived from their discrete counterparts) or by making assumptions that cannot be related to individual behavior. For instance, closure approximations (which assume independence between individuals at some stage) are commonly used, yet often generate errors in the resulting continuum model. In this work we use instead a systematic approach based on the method of matched asymptotic expansions to derive the macroscopic model of the two species system, generalizing the result \eqref{fpN_one}. 

The first part of this work is concerned with the derivation of the continuum PDE model from the microscopic particle-based model. We first introduce the microscopic description of the system based on $dN$-coupled stochastic differential equations (SDEs), where $d=2,3$ is the problem dimension, and its associated Fokker--Planck (FP) equation for the joint probability density of the system. We then perform a systematic asymptotic analysis of the FP equation in the limit of small but finite particle volume fraction, which results in a nonlinear cross-diffusion system of PDEs for the two subpopulations densities of each species. The nonlinear terms arise as a result of the excluded-volume effects in the system. We compare solutions to this model with stochastic simulations of the microscopic model to assess the validity of our approach, and find very good agreement. We also compare solutions for finite-size particles with those corresponding to point particles to investigate the importance of excluded-volume effects. 

Second, we examine how the continuum model can be used to determine the transport coefficients. The fact that the model keeps two distinct densities enables us to distinguish between two alternative notions of diffusion coefficient: the collective diffusion coefficient, which describes the evolution of the total density, and the self-diffusion coefficient, which describes the evolution of a single tagged particle.\cite{Hanna:1982gi} 
This is in contrast with the one-species model,\cite{Bruna:2012cg} from which we could only ``extract'' the collective diffusion coefficient. The reason for that is that, although both diffusion coefficients are defined for a system with identical particles, to describe the self-diffusion coefficient we need to \emph{tag} one individual particle and keep its probability density ``accessible'' in the continuum model. With the two species model this can easily be done by taking the tagged particle to constitute the second species. 
To our knowledge this is the first continuum PDE model that can be used to explain both types of diffusion. This makes it well suited to interpret experimental data from diffusion measurement experiments, which can often produce unexpected/misinterpreted results.\cite{Saunders:2012jl}

In the third part of this work, we explore the properties of the cross-diffusion PDE model. We find that rewriting the system in terms of its free energy and the mobility matrix can be a very useful tool to study the equilibria and stability of the system. In this alternative form, know as the gradient flow form, the evolution of the system can be interpreted as a probability flow down the gradients of the free energy. While this gradient flow structure is relatively well understood in the scalar case (one species), the task of obtaining an associated free energy  can be very challenging in the case of systems.\cite{Burger:2010gb}  For instance, in our case we are only able to write down an explicit free-energy functional under some conditions. Another use of the gradient-flow structure is that it gives an explicit upper bound on the particle volume fraction for the validity of our asymptotic model.  

\section{Two species model}
\label{sec:model}

Our starting point is a system of $N$ hard spheres (or disks) diffusing and interacting in a bounded domain $\Omega$ in $\mathbb R^d$ of typical dimensionless volume of order one. Suppose there are $N_b$ blue particles of diameter $\epsilon_b$ and constant diffusion coefficient $D_{b}$  and $N_r$ red particles of diameter $\epsilon_r$ and constant molecular diffusion coefficient $D_{r}$, with $N_b + N_r = N$. Note that we could have chosen to nondimensionalize so that one of the two dimensionless diffusion coefficients $D_b$ or $D_r$ is set to one. However, we choose deliberately not to do so so that the resulting model is symmetric upon exchange of the blue and red labels. Also note that we are not making explicit any relationship between the molecular diffusivity and the particle size (as it might exist if, say, the Stokes--Einstein relation holds in the system). 
We assume that the particles occupy a small volume fraction, so that
$N_b \epsilon_b ^d +  N_r \epsilon_r ^d\ll 1$.   
We suppose that the only interaction between particles is hard core repulsion (so that the particles cannot overlap), neglecting any electrostatic or hydrodynamic interaction forces. 

We denote the centers of the particles by ${\bf X} _i(t) \in \Omega$ at time $t\ge0$, where $1\le i \le N$. Each centre evolves according to the overdamped Langevin SDE
\begin{subequations}
\label{sde2}
\begin{align}
\label{sde_b}
&d {\bf{X}}_i (t) =    {\bf f}_b\bo ({\bf X} _i(t) \bo )\, dt + \sqrt{2D_b}\, d{\bf W}_i(t), \qquad 1\le i \le N_b, \\
\label{sde_r}
&d {\bf{X}}_i (t) =  {\bf f}_r \bo ({\bf X} _i(t) \bo )\, d t + \sqrt{2D_r}\, d{\bf W}_i(t), \qquad N_b+1\le i \le N, 
\end{align} 
\end{subequations}
where the ${\bf W}_i$ are $N$ independent $d$-dimensional standard Brownian motions and  ${\bf  f}_b$ and ${\bf  f}_r$  are the external forces on the blue and red particles, respectively. We restrict ourselves to the case where the external force acting on the $i$th particle depends only on its position ${\bf X} _i$, that is, ${\bf  f}_k \equiv {\bf  f}_k({\bf  X}_i): \Omega \to \mathbb R^d$. This excludes external forces such as electromagnetic and hydrodynamic forces, in which case ${\bf  f}_i$ would depend on the positions of all the particles $\vec X=  ({\bf  X}_1, \ldots,  {\bf  X}_N)$. We suppose that the initial positions ${\bf  X}_i(0)$ are  random and, within the same species, identically distributed.

Let  $P ({\vec x}, t)$ be the joint probability density function  
of the $N$ particles. Then, by the It\^o formula, $P (\vec x, t)$ 
evolves according to the  linear Fokker--Planck partial
differential equation (PDE) 
\begin{align}
\label{fp_junts}
\frac{\partial P}{\partial t} &= \vec \nabla_{\vec x} \cdot
\big[ \bo D \vec \nabla_{\vec x} \,  P -  \vec F(\vec x) \, P \big],
\end{align} 
where $\vec \nabla _{\vec x}$ and $\vec \nabla _{\vec x} \, \cdot$ respectively
stand for the gradient and divergence operators with respect to the 
$N$-particle position vector $\vec x  \in \Omega^{N}$. Here $\bo D=\text{diag} (D_b, \dots, D_b, D_r, \dots D_r)$ is the diffusivity matrix and $\vec F(\vec x) =  \bo ( {\bf f}_b ( {\bf x}_{1}), \dots,  {\bf f}_b ( {\bf x}_{N_b}),\allowbreak {\bf f}_r ( {\bf x}_{N_b+1}), \dots,  {\bf f}_r ( {\bf x}_{N}) \bo )$ is the $dN$-dimensional drift. 
Splitting the position vector $\vec x$ into the position vector for the blue particles, $\vec x_b = ({\bf x}_1,\dots,  {\bf x}_{N_b})$, and for the red particles, $\vec x_r = ({\bf x}_{N_b + 1},\dots,  {\bf x}_{N})$, equation \eqref{fp_junts} can be rewritten as 
\begin{subequations}
\label{twofp}
\begin{align}
\label{twofp_eq}
\frac{\partial P}{\partial t} =   \vec \nabla_{\vec x_b} \cdot
\big[ D_b \vec \nabla_{\vec x_b} \,  P -  \vec F_b(\vec x_b) \, P \big] + \vec \nabla_{\vec x_r}  \cdot  \big[D_r \vec \nabla_{\vec x_r} \,  P -  \vec F_r(\vec x_r) \, P \big] \quad  \textrm{in} \quad  \Omega_\epsilon^{N},
\end{align}
where $\vec F_b(\vec x_b)$ is the drift acting on the blue particles [first $dN_b$ components of $\vec F(\vec x)$], and analogously for $\vec F_r(\vec x_r)$. Because of excluded-volume effects, \eqref{twofp_eq} is not defined in $\Omega ^N$ but in its ``hollow form''   $\Omega_\epsilon ^N = \Omega ^N \setminus \mathcal B_\epsilon$, where $\mathcal B_\epsilon$ is the set of all illegal configurations (with at least one overlap), 
\begin{equation*}
\mathcal B_\epsilon=\left \{\vec x\in \Omega^N: \exists i\ne j \quad \textrm{
s.t. } \quad  \| {\bf x}_i -  {\bf x}_j \| \le \tfrac{1}{2}(\epsilon_i + \epsilon_j)
\right \},
\end{equation*}
where $\epsilon_i = \epsilon_b$ for $i\le N_b$ and $\epsilon_i = \epsilon_r$ otherwise. The domain of definition $\Omega_\epsilon^N$ is known as the \emph{configuration space}. On the collision surfaces $\partial \Omega_\epsilon ^N $ we have the reflecting boundary condition
\begin{equation} 
\label{twofp_bc} 
\big[ \vec \nabla_{\vec x} \,  P -  \vec F(\vec x) \, P \big] \cdot  {\vec n} = 0 \qquad \text{on} \qquad  \partial \Omega_\epsilon ^N, 
\end{equation}
where $ {\vec n}  \in \mathcal S^{dN-1}$ denotes the unit outward normal. The initial condition on $P$ is
\begin{equation}
P(\vec x, 0 ) = P_0 (\vec x).
\end{equation}
\end{subequations}
Since all the particles within the same species (blue or red) are identically distributed, the initial distribution $P_0(\vec x)$ is invariant to permutations of the particles labels \emph{within the same species}. The form of \eqref{twofp} then means that $P$ itself is invariant to permutations of the blue or red particle labels for all time. 

As in Ref.~\onlinecite{Bruna:2012cg}, our goal is to reduce the high-dimensional PDE model \eqref{twofp} to a low dimensional PDE model for the marginal density function of one particle. However, as mentioned in the introduction, instead of obtaining an equation for $p({\bf  x}_1,t) =  \int P(\vec x,t) \  \mathrm{d} {\bf  x}_{2} \cdots \mathrm{d} {\bf  x}_{N}$ comprising the collective effect of $N$ identical particles, now we will have two marginal density functions, one representative of the blue particles and one representative of the red particles. Because all blue particles are identical and all red particles are identical, we are interested in the marginal density function of, say, the first blue particle and the last red particle, given by 
\begin{subequations}
\begin{align*}
b(  {\bf x},t) = \int_{\Omega_\epsilon^N} P ({\vec x}, t) \, \delta ({\bf x} - {\bf x}_1)\, \ud \vec x, \hspace{1.5cm} r(  {\bf x},t) = \int_{\Omega_\epsilon^N} P ({\vec x}, t) \, \delta ({\bf x} - {\bf x}_N) \,\ud \vec x,
\end{align*}
\end{subequations}
respectively. 
We aim to reduce the high-dimensional PDE for $P$ \eqref{twofp} to a low-dimensional \emph{system} of PDEs for $b$ and $r$ through a systematic asymptotic expansion as $\epsilon_b, \epsilon_r \to 0$.

\subsection{Point particles}

In the particular case of point particles ($\epsilon_b = \epsilon_r = 0$) the model reduction is straightforward. In this case the $N$ particles are
independent and the domain is $\Omega_\epsilon ^N \equiv \Omega ^N$
(no holes), which implies that the internal boundary conditions in
\eqref{twofp_bc} vanish. Therefore 
\begin{equation}
\label{Ppoints}
P(\vec x,t) = \prod_{i=1}^{N_b} b(  {\bf x}_i,t) \prod_{i=N_b+1}^{N} r(  {\bf x}_i,t),
\end{equation}
and the evolution equations for the one-particle density functions $b$ and $r$ follow from integrating equation \eqref{twofp_eq} multiplied by $\delta ({\bf x} - {\bf x}_1)$ and $\delta ({\bf x} - {\bf x}_N)$, respectively, over the configuration space $\Omega^N$ using \eqref{Ppoints}
\begin{subequations}
\label{2:point}
\begin{align}
\label{point_eqb}
\frac{\partial b}{\partial t}({\bf  x},t) &=  \bo \nabla_{
{\bf x}} \cdot \left[ D_b\bo \nabla_{{\bf  x}} \,  b -  
{\bf f}_b({\bf  x}) \, b \right],\\ 
\label{point_eqr}
\frac{\partial r}{\partial t}({\bf  x},t) &=  \bo \nabla_{
{\bf x}} \cdot \left[D_r \bo \nabla_{{\bf  x}} \,  r -  
{\bf f}_r({\bf  x}) \, r \right],
\end{align}
in $\Omega \times \mathbb R^+$. The boundary conditions \eqref{twofp_bc} become  simple no-flux boundary conditions on the domain walls,
\begin{align}
\label{point_bcb} 
0&= \left[ D_b\bo \nabla_{{\bf  x}_1} \,  b -  {\bf  f}_b(
{\bf x}_1) \, b \right] \cdot \boldsymbol { \hat {\bf n}}_1,
\\
\label{point_bcr} 
0&= \left[ D_r\bo \nabla_{{\bf  x}_1} \,  r -  {\bf  f}_r(
{\bf x}_1) \, r \right] \cdot \boldsymbol { \hat {\bf n}}_1,
\end{align}
on $\partial \Omega \times \mathbb R^+$, where $\boldsymbol {\hat  {\bf n}}_1$ is the outward unit normal to $\partial \Omega$. The system is supplemented by initial values
\begin{equation}
\label{point_initial}
b({\bf  x},0) = b_0({\bf  x}), \hspace{1.5cm} r({\bf  x},0) = r_0({\bf  x}),
\end{equation}
in $\Omega$. Here $b_0 ({\bf x}) = \int_{\Omega^N} P_0 ({\vec x}, t) \, \delta ({\bf x} - {\bf x}_1)\, \ud \vec x$, and similarly for $r_0$.
\end{subequations}

\subsection{Finite-size particles}

When $\epsilon_b$ and/or $\epsilon_r$ are greater than zero, the internal boundary conditions in \eqref{twofp_bc} mean the particles are no longer independent. The integration of \eqref{twofp} over ${\bf x}_2, \dots, {\bf x}_N$ results in integrals over the \emph{collision surfaces}, on which $P$ must be evaluated. When the particle volume fraction is small, the dominant contributions to these  collision integrals correspond to two-particle collisions. This implies that if the case of $N=2$ can be solved it can easily extended to arbitrary $N$.\cite{Bruna:2012cg} For two species three types of interaction with $N=2$ are possible: either two blue particles, either two red particles, or one blue particle and one red particle interacting. We note that the first two types involving two \emph{identical} hard spheres have already been computed in Ref.~\onlinecite{Bruna:2012cg} yielding equation \eqref{fpN_one}. Therefore, here we only need to consider the third case, that is, $N_b = N_r = 1$.

For one blue particle at position ${\bf x}_1$ and one red particle at position ${\bf x}_r$, equation  \eqref{twofp_eq} reads
\begin{subequations}
\label{twocolors}
\begin{align}
\label{twocolors_eq}
&\frac{\partial P}{\partial t}( {\bf x}_1,  {\bf x}_2, t)
= \bo \nabla_ { {\bf x}_1} \cdot \left[D_b \bo \nabla_{
{\bf x}_1} \,  P -   {\bf f}_b( {\bf x}_1) \, P
\right]  +   \bo \nabla_ { {\bf x}_2} \cdot \left[D_r \bo \nabla_ { {\bf
x}_2} \,  P -   {\bf f}_r( {\bf x}_2) \, P \right],
\end{align}
for  $({\bf  x}_1, {\bf  x}_2) \in \Omega_\epsilon ^2$,
and the boundary condition \eqref{twofp_bc} reads 
\begin{equation}
\label{twocolors_bc}
\left[ D_b \bo \nabla_{ {\bf x}_1} \,  P -   {\bf f}_b({\bf x}_1) \, P \right] \cdot  {\hat {\bf n}}_1 +  \left[D_r \bo \nabla_ { {\bf x}_2} \,  P -   {\bf f}_r( {\bf x}_2) \, P \right] \cdot  \hat{{\bf n}}_2 = 0, 
\end{equation}
\end{subequations}
on ${\bf  x}_i \in \partial \Omega$ and $\|{\bf  x}_1 -
{\bf  x}_2\| = \tfrac{1}{2}(\epsilon_b + \epsilon_r)$. Here $ \boldsymbol {\hat {\bf n}}_i =  { {\bf n}}_i / \| { {\bf n}}_i \|$, where $ { {\bf n}}_i$ is the
component of the normal vector $\vec  n$ corresponding to the $i$th
particle, $\vec n = ({\bf  n}_1, {\bf  n}_2)$. We note
that $\boldsymbol  {\hat {\bf n}}_1 = 0$ on ${\bf  x}_2 \in \partial
\Omega$, and that $\boldsymbol {\hat {\bf n}}_1 = -\boldsymbol {\hat {\bf n }}_2$ on $\|
{\bf x}_1-{\bf  x}_2\| = \epsilon$.    
It is convenient to introduce $\epsilon_{br}$ as the distance at contact between one blue particle and one red particle, $\epsilon_{br}=(\epsilon_b + \epsilon_r)/2$.

We proceed to obtain an equation for $b({\bf x},t)$ from \eqref{twocolors}. 
We denote by  $\Omega ({\bf  x}_1)$ the region available to particle 2 (the red particle) when particle 1 (the blue particle) is at ${\bf x}_1$, namely, $\Omega({\bf  x}_1) = \Omega \setminus B_{\epsilon_{br}} ({\bf  x}_1)$. Since the domain dimensions are much larger than the particle diameters, the volume $|\Omega({\bf  x}_1)|$ is constant  to leading order. Integrating Eq. \eqref{twocolors_eq} over $\Omega ({\bf  x}_1)$ 
yields
\begin{align}
\label{twofp2r} 
\begin{aligned}
\frac{\partial b}{\partial t}({\bf  x}_1, t) =  \mbox{ } & \bo \nabla_{{\bf  x}_1} \cdot \left[ D_b\bo \nabla_{{\bf  x}_1} \,b  -  {\bf  f}_b({\bf  x}_1) \, b \right]
\\ 
&+ \int_{\partial B_{\epsilon_{br}} ({\bf  x}_1 )} [{\bf f}_b({\bf  x}_1) \, P -2D_b \bo \nabla_{ {\bf x}_1} P -D_b \bo \nabla_{ {\bf  x}_2} P ]\cdot  \hat {\bf n}_2 \, \mathrm{d} S_2 \qquad 
\\  
&+ \int_{\partial \Omega\cup\partial B_{\epsilon_{br}} ({\bf  x}_1 )}  \left[D_r \bo \nabla_{{\bf  x}_2} \,  P -
{\bf  f}_r({\bf  x}_2) \, P \right] \cdot  {\boldsymbol
{\hat {\bf  n}}}_2 \, \mathrm{d} S_2.  
\end{aligned}
\end{align} 
The first  integral  in (\ref{twofp2r}) comes from switching the order of integration with respect to ${\bf x}_2$ and differentiation with respect to ${\bf x}_1$ using the transport theorem; the second  comes from using the divergence theorem on the derivatives in ${\bf x}_2$. Using \eqref{twocolors_bc} and rearranging we find 
\begin{align}
\label{twofp2rr}
\begin{aligned}
\frac{\partial b}{\partial t}({\bf  x}_1, t)  &=  
\bo \nabla_{{\bf  x}_1} \cdot \left[D_b \bo \nabla_{{\bf  x}_1} \, b -  {\bf  f}_b({\bf  x}_1) \, b \right] -2 D_b  \! \int_{\partial B_{\epsilon_{br}} ({\bf  x}_1 )}  \! \! \bo \nabla_{ {\bf x}_1} P \cdot  \hat  {\bf n}_2 \, \mathrm{d} S_2 
\\
 & \quad  + \!  \int_{\partial B_{\epsilon_{br}} ({\bf  x}_1 )} \! \!  \left \{   (D_r-D_b) \bo \nabla_{ {\bf x}_1} P + P \left[  {\bf f}_b({\bf  x}_1) -  {\bf  f}_r({\bf  x}_2) \right]\right \} \cdot  \hat  {\bf n}_2 \, \mathrm{d} S_2 
\\
& =  \bo \nabla_ { {\bf x}_1} \cdot \left[D_b \bo \nabla_ { {\bf x}_1} b -   {\bf f}_b( {\bf x}_1) b  \right] - D_b \! \int_{\partial B_{\epsilon_{br}} ({\bf x}_1 )} \! \! \left(\bo \nabla_ { {\bf x}_1} P + \bo \nabla_ { {\bf x}_2} P \right) \cdot  { \hat {\bf n}}_2 \, \ud S_2.
\end{aligned}
\end{align}
We denote the integral above by
\begin{equation}
\label{Ibr}
\mathcal I_{br} ({\bf x}_1) =  - D_b\int_{\partial B_{\epsilon_{br}} ({\bf x}_1 )} \left(\bo \nabla_ { {\bf x}_1} P + \bo \nabla_ { {\bf x}_2} P \right) \cdot  { \hat {\bf n}}_2 \, \ud S_2.
\end{equation}
At this stage, it is common to use a closure approximation.\cite{Felderhof:1978vn,Beenakker:1983wr} For example, the classical closure approximation is to assume that particles are not correlated, that is, $P({\bf x}_1,{\bf x}_2,t) = b({\bf x}_1,t) r({\bf x}_2,t)$. However, the pairwise particle interaction---and therefore the correlation between their positions---is \emph{exactly} localized near the collision surface  $\partial B_{\epsilon_{br}}({\bf x}_1)$. Here we use a systematic alternative method based matched asymptotic expansions \cite{Holmes:1995uo} to compute $\mathcal I_{br} ({\bf x}_1)$.

\subsection{Matched asymptotic expansions of $P$}
\label{sec:2_MAE}

We suppose that when two particles are far apart ($\|{\bf  x}_1\! -{\bf  x}_2\|\gg \epsilon_{br}$) they are independent, whereas when they are close to each other ($\|{\bf  x}_1 - {\bf  x}_2\| \sim \epsilon_{br}$) they are correlated. We designate these two regions of configuration space the outer region and inner region, respectively.

In the outer region we define $P_{out}( {\bf x}_1,  {\bf x}_2 , t)= P( {\bf x}_1,  {\bf x}_2 , t)$. By independence, we have that
\begin{equation} 
\label{indep}
P_{out}({\bf  x}_1, {\bf  x}_2, t) = q_b({\bf  x}_1, t) q_r({\bf  x}_2, t), 
\end{equation}
for some functions $q_b({\bf  x}, t)$ and $q_r({\bf  x}, t)$. It is important to note that these functions will not be the same in general since $P$ is not invariant to a switch of blue and red particle labels. Also note that we could introduce more terms in the asymptotic expansion for $P_{out}$. However, the subsequent analysis shows that the value of the integral $\mathcal I_{br}$ is invariant to the first-order correction of $P_{out}$ which we thus not need to consider further.

In the inner region, we set  ${\bf  x}_1 = \tilde {\bf x}_1$ and ${\bf  x}_2 =\tilde {\bf x}_1 + \epsilon_{br} \tilde {\bf x}$ and define $\tilde P (\tilde {\bf x}_1, \tilde {\bf x}, t) = P( {\bf x}_1,  {\bf x}_2 , t)$. Inserting this change of variables into \eqref{twocolors} yields
\begin{subequations}
\label{innertwo}
\begin{align}
\label{innertwo_eq} 
\begin{aligned}
\epsilon_{br}^2 \frac{\partial \tilde P}{\partial t} (\tilde {\bo {\mathrm x}}_1, \tilde {\bo {\mathrm x}}, t)  =  &  \  (D_b + D_r) \bo \nabla_{\tilde { {\bo {\mathrm x}}}}^2 \, \tilde P  
\\
&+ \epsilon_{br}  \bo \nabla _{\tilde{\bo {\mathrm x}}} \cdot \big\{ \left[{\bf  f}_b(\tilde {\bo {\mathrm x}}_1) -  {\bf  f}_r(\tilde {\bo {\mathrm x}}_1 + \epsilon_{br}   \tilde {\bo {\mathrm x}}) \right] \tilde P -2 D_b   \bo \nabla_{\tilde {\bo {\mathrm x}}_1} \tilde P \big \} \\
&+  \epsilon_{br}^2 \big\{ D_b  \bo \nabla_{\tilde {\bo {\mathrm x}}_1}^2 \, \tilde P  -\bo \nabla _{\tilde{\bo {\mathrm x}}_1} \cdot \big[ {\bf  f}_b(\tilde {\bo {\mathrm x}}_1)  \tilde P \big] \big\}, 
\end{aligned}
\end{align}
with
\begin{align}
\label{innertwo_bc}
\tilde {\bo {\mathrm x}} \cdot  \bo \nabla_{\tilde {\bo {\mathrm x}}} \tilde P =   \frac{\epsilon_{br}}{D_b + D_r} \, \tilde {\bo {\mathrm x}} \cdot \big \{
D_b\bo \nabla_{\tilde {\bf  x}_1} \tilde P + \left[  {\bf 
f}_r(\tilde {\bo {\mathrm x}}_1 +\epsilon_{br} \tilde {\bo {\mathrm x}})
- {\bf  f}_b(\tilde {\bo {\mathrm x}}_1) \right] \tilde P,
\big\}
\end{align}
on $\| \tilde {\bo {\mathrm x}} \| = 1$. Note that now \eqref{innertwo_bc} contains the no-flux boundary condition on the contact between the two particles, but not on $\partial \Omega$. As pointed out above, this is because we can assume that  $\tilde {\bf x}_1$ is not close to $\partial \Omega$; the boundary  only affects higher-order terms.   
In addition to \eqref{innertwo_bc} the inner solution must match with
the outer solution $P_{out}$ as $\|\tilde {\bf x}\| \to \infty$.  Expanding \eqref{indep} in inner variables gives
\begin{align}
\label{2:bc_match}
\begin{aligned}
\tilde P ( \tilde {\bf  x}_1, \tilde {\bf x}, t) & \to q_b(\tilde {{\bf 
x}}_1, t) q_r(\tilde {\bf x}_1 + \epsilon_{br} \tilde {
{\bf  x}}) \\
& \sim  q_b (\tilde {\bf x}_1,t) q_r (\tilde {\bf x}_1,t) +  \epsilon_{br} q_b(\tilde {
{\bf  x}}_1) \, \tilde {\bf x} \cdot \bo \nabla _{ \tilde
{\bf x}_1} q_r( \tilde {\bf x}_1) + \cdots,
\end{aligned}
\end{align}
\end{subequations}
as $\| \tilde {\bf x} \| \to \infty$. 
We now look for a solution to \eqref{innertwo} of the form
$\tilde P (\tilde {\bo {\mathrm x}}_1, \tilde {\bo {\mathrm x}}, t) \sim \tilde P^{(0)} (\tilde {\bo {\mathrm x}}_1, \tilde {\bo {\mathrm x}}, t) + \epsilon_{br}\tilde P^{(1)} (\tilde {\bo {\mathrm x}}_1, \tilde {\bo {\mathrm x}}, t)  +  \cdots$. 
The leading-order inner problem is simply 
\begin{alignat}{3}
\begin{aligned}
\bo \nabla_{\tilde {\bo {\mathrm x}}}^2 \tilde P^{(0)} &= 0, & & &  &  \\
\tilde {\bo {\mathrm x}} \cdot  \bo \nabla _{\tilde {\bo{\mathrm x}}} \tilde P^{(0)} &= 0& \qquad  &\textrm{on}& \qquad \|\tilde {\bo {\mathrm x}}\| &= 1,\\
\tilde P^{(0)} &\sim q_b (\tilde {\bo {\mathrm x}}_1,t) q_r (\tilde {\bo {\mathrm x}}_1,t)  & & \textrm{as} & \|\tilde {\bo {\mathrm x}}\| &\to \infty,
\end{aligned}
\end{alignat}
which is trivially solved by
\begin{equation}
\label{two_order0_sol}
\tilde P^{(0)}(\tilde {\bo {\mathrm x}}_1, \tilde {\bo {\mathrm x}},t)  =  q_b (\tilde {\bo {\mathrm x}}_1,t) q_r (\tilde {\bo {\mathrm x}}_1,t).
\end{equation} 
At $\mathcal O(\epsilon_{br})$ \eqref{innertwo} reads, using \eqref{two_order0_sol} and Taylor-expanding ${\bf f}_b$ and ${\bf f}_r$, 
\begin{alignat}{3}
\label{two_order1}
\begin{aligned}
\bo \nabla_{\tilde {  {\bf  x}}}^2 \, \tilde P^{(1)} &= 0, & & & &
\\
\tilde {\bo {\mathrm x}} \cdot  \bo \nabla _{\tilde {\bo{\mathrm x}}} \tilde P^{(1)} &=  \tilde {\bo {\mathrm x}} \cdot {\bf A}(\tilde {\bo {\mathrm x}}_1,t), & \qquad &  \textrm{on} & \qquad \|\tilde {\bo {\mathrm x}}\| &= 1,\\
\tilde P^{(1)} &\sim  \tilde {\bo {\mathrm x}} \cdot {\bf B}(\tilde {\bo {\mathrm x}}_1,t), & & \textrm{as}&  \|\tilde {\bo {\mathrm x}}\| &\to \infty,
\end{aligned}
\end{alignat}
with
\begin{align}
\label{AandB}
\begin{aligned}
{\bf A}(\tilde {\bo {\mathrm x}}_1,t) &=   \frac{1}{D_b + D_r} \big \{
D_b\bo \nabla_{\tilde {\bo {\mathrm x}}_1} (q_b \, q_r) + \left[  {\bf  f}_r(\tilde {\bo {\mathrm x}}_1)  - {\bf  f}_b(\tilde {\bo {\mathrm x}}_1) \right] q_b \, q_r \big\}, \\
{\bf B}(\tilde {\bo {\mathrm x}}_1,t) &= q_b \bo \nabla _{ \tilde {\bo {\mathrm x}}_1} q_r.
\end{aligned}
\end{align}
The solution to \eqref{two_order1} is
\begin{equation}
\label{two_order1_sol}
\tilde P^{(1)}(\tilde {\bo {\mathrm x}}_1, \tilde {\bo {\mathrm x}},t)  =  \tilde {\bo {\mathrm x}} \cdot {\bf A}(\tilde {\bo {\mathrm x}}_1,t) + \left (  \tilde {\bo {\mathrm x}} + \frac{ \tilde {\bo {\mathrm x}}}{ (d-1) \| \tilde {\bo {\mathrm x}}\|^d}\right ) \cdot \left [ {\bf B}(\tilde {\bo {\mathrm x}}_1,t)  - {\bf A}(\tilde {\bo {\mathrm x}}_1,t) \right].
\end{equation}
Combining \eqref{two_order0_sol} and \eqref{two_order1_sol}, the inner solution is
\begin{align}
\label{2:sol_inner}
\begin{aligned}
\tilde P (\tilde {\bf x}_1, \tilde {\bf x}, t)
\sim \ & q_b  q_r + \epsilon_{br}  \, q_b \, \tilde {\bo {\mathrm x}}  \cdot  \bo \nabla _{ \tilde {\bo {\mathrm x}}_1} q_r  +  \frac{ \epsilon_{br} }{(D_b + D_r) (d-1) \| \tilde {\bo {\mathrm x}}\|^d}  \\
&  \times \tilde {\bo {\mathrm x}} \cdot \Big \{ [  {\bf  f}_b(\tilde {\bo {\mathrm x}}_1)   -  {\bf  f}_r(\tilde {\bo {\mathrm x}}_1) ] q_b \, q_r + D_r q_b \bo \nabla _{ \tilde {\bo {\mathrm x}}_1} q_r - D_b q_r \bo \nabla _{ \tilde {\bo {\mathrm x}}_1} q_b\Big\} + \mathcal O(\epsilon_{br}^2),
\end{aligned}
\end{align}
where $q_b = q_b (\tilde {\bo {\mathrm x}}_1,t)$ and $q_r = q_r (\tilde {\bo {\mathrm x}}_1,t)$. 
Comparing the inner solution \eqref{2:sol_inner} in the case of two distinguishable particles with that of two identical particles, for which $\tilde P (\tilde {\bf x}_1, \tilde {\bf x}, t) \sim   q^2 + \epsilon  q \, \tilde {\bo {\mathrm x}}  \cdot  \bo \nabla _{ \tilde {\bo {\mathrm x}}_1} q$, we find that two extra terms contribute to the inner solution now: one is due to the difference in the drift force acting on each particle, $[ {\bf  f}_b(\tilde {\bo {\mathrm x}}_1)   -  {\bf  f}_r(\tilde {\bo {\mathrm x}}_1) ] \, q_b  q_r$, and the other owing to the different initial conditions and/or diffusivities, $D_r q_b \bo \nabla _{ \tilde {\bo {\mathrm x}}_1} q_r - D_b q_r \bo \nabla _{ \tilde {\bo {\mathrm x}}_1} q_b$. Note however that, even in the case of (physically) identical particles, we could use the two-species distinction to have the particles subdivided into two groups, each group with different initial conditions. In that case, all the physical parameters $\epsilon_i$, $D_i$ and ${\bf f}_i$ are identical, and the distinction between the two groups in the model is contained in the outer functions $q_b$ and $q_r$.

Using the inner solution \eqref{2:sol_inner},  we can now evaluate the integral $\mathcal I_{br}({\bf x}_1)$ in \eqref{Ibr}.  Expressing it in terms of the inner variables, we obtain 
\begin{align}
\label{intd}
\begin{aligned}
\mathcal I_{br}(\tilde {\bf x}_1) \sim  \epsilon_{br} ^d  \frac{D_b}{D_b + D_r} \frac{2\pi}{d}  \bo \nabla_{ \tilde {\bf x}_1}  \cdot  \Big \{ & \big[(d\!-\!1)D_b + dD_r \big]q_b \bo \nabla _{ \tilde { {\bf x}}_1} q_r  - D_b q_r \bo \nabla_{\tilde {\bo {\mathrm x}}_1} q_b \\
& + \big[  {\bf  f}_b(\tilde { {\bf x}}_1)  - {\bf  f}_r(\tilde {{\bf x}}_1) \big] q_b \, q_r \! \Big \}+\cdots 
\end{aligned}
\end{align}
Now we use the normalization condition on $P$ to determine the outer functions $q_b$ and $q_r$. We find that $q_b({\bf x}) = b ({\bf x}) + \mathcal O(\epsilon_{br}^d)$ and $ q_r({\bf x}) = r ({\bf x}) + \mathcal O(\epsilon_{br}^d)$, which will allow us to express $\mathcal I_{br}({\bf x}_1)$ in terms of the densities $b$ and $r$. 

\subsection{System of equations for $b$ and $r$ }
\label{sec:final_model}

Inserting the expression for $\mathcal I_{br}({\bf x}_1)$ obtained in \eqref{intd} into \eqref{twofp2rr} we find, to $\mathcal O(\epsilon_{br}^d)$,
\begin{align}
\label{N2blue}
\begin{aligned}
\frac{\partial {b}}{\partial t} ( {\bf x}_1, t)  =  
\bo \nabla_{{\bf x}_1} \cdot  \Big( &  D_b \bo \nabla_{{\bf x}_1 }
{ b} - {\bf f}_b( {\bf x}_1)  b \\ &+ \epsilon_{br}^d D_b \big \{\beta_b \, {b} \bo \nabla_{{\bf x}_1} {r} - \gamma_b {r}\bo \nabla_{{\bf x}_1} {b}\big \} +  \epsilon_{br}^d \gamma_b \left[  {\bf f}_b( {\bf x}_1) -  {\bf f}_r( {\bf x}_1) \right] br \Big),   
\end{aligned}
\end{align}
where  
\begin{equation*}
\begin{aligned}
\beta_b =  \frac{2\pi}{d}  \frac{  [(d-1)D_b  +  dD_r]}{D_b + D_r}, \qquad 
\gamma_b =  \frac{2\pi}{d} \frac{D_b}{D_b + D_r}.
\end{aligned}
\end{equation*}
Equation \eqref{N2blue} gives the evolution of $b$ for a system with one blue particle and one red particle. The extension from one to $N_r$ red particles is straightforward up to $\mathcal O(\epsilon_{br}^d)$, since at this order only pairwise interactions need to be considered. For $N_r$ arbitrary, the blue particle has $N_r$ blue--red inner regions, one with each of the $N_r$ red particles; hence there are $N_r$ copies of the  nonlinear term in \eqref{N2blue}. Similarly, for $N_b$ arbitrary the blue particle can have blue--blue pairwise interactions with any of the $N_b -1$ remaining blue particles; the contribution of a pairwise interaction between identical particles is found in \eqref{fpN_one}. Thus the blue marginal density function satisfies
\begin{subequations}
\label{finalmodel_compact}
\begin{align}
\label{finalmodelc_b}
\begin{aligned}
\frac{\partial {b}}{\partial t} ( {\bf x}, t)  =  
\bo \nabla_{{\bf x}}  \cdot \Big( &  D_b \bo \nabla_{{\bf x}} 
{ b} - {\bf f}_b( {\bf x})  b   +
(N_b-1)  \epsilon_b^d D_b \alpha    b  \bo \nabla_{{\bf x}} { b}  \\ 
& +  N_r \epsilon_{br}^d \big \{ D_b (\beta_b \, {b} \bo \nabla_{{\bf x}} {r} - \gamma_b {r}\bo \nabla_{{\bf x}} {b} ) + \gamma_b \left[  {\bf f}_b( {\bf x}) -  {\bf f}_r( {\bf x}) \right] br\big \} \Big),
\end{aligned}
\end{align}
in $\Omega\times \mathbb R^+$. A similar procedure shows that the red marginal density $r$ satisfies
\begin{align}
\label{finalmodelc_r}
\begin{aligned}
\frac{\partial {r}}{\partial t} ( {\bf x}, t) =    
\bo \nabla_{{\bf x}}  \cdot  \Big(& D_r \bo \nabla_{{\bf x}} 
{ r} -  {\bf f}_r( {\bf x})  r + (N_r-1)  \epsilon_r^d D_r \alpha    r  \bo \nabla_{{\bf x}} { r}   \\
&+ N_b \epsilon_{br}^d  \big \{ D_r(\beta_r \, {r} \bo \nabla_{{\bf x}} {b} - \gamma_r b\bo \nabla_{{\bf x}} r) + \gamma_r \left[  {\bf f}_r( {\bf x}) -  {\bf f}_b( {\bf x}) \right] br \big \} \Big), 
\end{aligned}
\end{align}
in $\Omega\times \mathbb R^+$. This system is complemented with no-flux boundary conditions on $\partial \Omega\times \mathbb R^+$ and initial conditions
\begin{equation}
\label{finalmodelc_ini}
b({\bf  x},0) = b_0({\bf  x}), \hspace{1.5cm} r({\bf  x},0) = r_0({\bf  x})
\end{equation}
in $\Omega$. The coefficients in the equations ($i=b$ and $j=r$ and vice versa) are as follows
\begin{equation}
\label{coef_23d}
\begin{aligned}
\alpha &= \frac{2(d-1)\pi}{d}, \qquad   \beta_i =    \frac{2\pi}{d}  \frac{ [(d-1)D_i  +  dD_j]}{D_i + D_j}  ,  \qquad \gamma_i=  \frac{2\pi}{d} \frac{D_i}{D_i + D_j}   ,
\end{aligned}
\end{equation}
for $d=2$ or 3.
\end{subequations}

We have obtained a nonlinear cross-diffusion system for the blue and red marginal probability densities, which captures the enhancement (diminishment) of the effective diffusion rate, due to excluded-volume interactions between particles of the
same species (of the other species). Namely, the diffusion of one blue particle is enhanced by collisions with the other blues, and reduced by collisions with red particles [terms with $+\alpha  b  \bo \nabla_{{\bf x}} { b} $ and $-\gamma_b  r  \bo \nabla_{{\bf x}} { b}$, respectively, in \eqref{finalmodelc_b}]. This will allows us to distinguish between two alternative notions of diffusion coefficient: the collective (or mutual) diffusion coefficient and the self-diffusion coefficient (see \secref{sec:diffusions}). Note that the reduced model \eqref{finalmodel_compact} has now a nonlinear drift term. This is effectively a ``drag'' term due to the different drift velocities of the red and blue particles.

It is reassuring to find that we can recover the system for a single species from the system for two species \eqref{finalmodel_compact}. When the two species are identical, that is, $D_b = D_r$, $\epsilon_b = \epsilon_r$, and ${\bf f}_b = {\bf f}_r $, the governing equations \eqref{finalmodelc_b} and \eqref{finalmodelc_r} of the densities $b$ and $r$ are the same. Then, if the initial densities \eqref{finalmodelc_ini} are equal, $b_0({\bf x})=r_0({\bf x}) := p_0({\bf x})$, then $b({\bf x}, t)=r({\bf x}, t) := p({\bf x}, t)$ for all times. Consequently, we can replace $b$ and $r$ by $p$ in \eqref{finalmodelc_b} and recover the one-species equation \eqref{fpN_one}, by noting that, when $D_b = D_r$, $\beta_b - \gamma_b$ is equal to $\alpha$ [see \eqref{coef_23d}]. In this scenario, the nonlinear terms in \eqref{finalmodelc_b} rearrange to 
\begin{align*}
(N_b -1) \epsilon^d D_b \alpha p \bo \nabla_{\bf x} p + N_r \epsilon^d D_b(\beta_b - \gamma_b)  p \bo \nabla_{\bf x} p \equiv (N_b + N_r -1) \epsilon^d D_b \alpha  p \bo \nabla_{\bf x} p,
\end{align*}
which coincides with the nonlinear term in the one-species equation \eqref{fpN_one}. 

Finally, note that we have not specified any relation between the diffusion coefficients of a single particle ($D_b$ and $D_r$) and the particles' diameters ($\epsilon_b$ and $\epsilon_r$), even though
 a relation may exist between these parameters. For instance, if we are modeling a system for which the Stokes--Einstein relation holds, then the diffusivity should be inversely proportional to the particle's diameter. Thus it may be that not all of the four parameters  $D_b$, $D_r$, $\epsilon_b$, and $\epsilon_r$ can be chosen independently.

\subsection{System in matrix form} \label{sec:diff_matrix_form}

The system \eqref{finalmodel_compact} may be rewritten in the form 
\begin{subequations}
\label{model_matrix}
\begin{equation}
\label{matrix_eq}
\frac{\partial}{\partial t} \begin{pmatrix} b \\ r \end{pmatrix} ({\bf x},t) = \bo \nabla_{\bf x} \cdot \left[ {\bf D}(b,r) \bo \nabla_{\bf x} \begin{pmatrix} b \\ r \end{pmatrix} - {\bf F}(b,r) \begin{pmatrix} b \\ r \end{pmatrix}
\right],
\end{equation} 
where ${\bf D}(b,r) = \left(\begin{smallmatrix} D_{bb} & D_{br} \\ D_{rb} & D_{rr} \end{smallmatrix} \right)$ is the diffusion matrix,
\begin{equation}
\label{diffusionmatrix}
\renewcommand{\arraystretch}{1.1}
\textstyle
{\bf D}(b,r) = 
\begin{pmatrix} D_{b}\left[ 1 + (N_b\!-\!1) \epsilon_b^d   \alpha b - N_r\epsilon_{br}^d \gamma_b  r\right]  & D_b N_r \epsilon_{br}^d  \beta_b  b \\ D_r N_b \epsilon_{br}^d  \beta_r  r & D_{r}\left[ 1 + (N_r\!-\!1) \epsilon_r^d   \alpha r - N_b \epsilon_{br}^d \gamma_r  b \right] \end{pmatrix},
\end{equation}
and ${\bf F}(b,r)$ is the drift matrix
\begin{equation}
\renewcommand{\arraystretch}{1.1}
{\bf F}(b,r) = \begin{pmatrix} {\bf f}_b ({\bf x} ) & N_r \epsilon_{br}^d \gamma_b[  {\bf f}_r( {\bf x}) -  {\bf f}_b( {\bf x})] b \\
N_b\epsilon_{br}^d  \gamma_r[  {\bf f}_b( {\bf x}) -  {\bf f}_r( {\bf x})] r & {\bf f}_r ({\bf x} )
\end{pmatrix},
\end{equation}
\end{subequations}
with the coefficients given in \eqref{coef_23d}. 
In order to focus on the diffusion part of \eqref{matrix_eq}, we set ${\bf F}\equiv 0$ for the rest of this section. 

An important consideration is that the reduced continuum model we have obtained for the collective or population-level behavior is different to the corresponding continuum limit of the discrete on-lattice counterpart model:\cite{Burger:2010gb} the coefficients of our advection--diffusion system \eqref{model_matrix} derived from the off-lattice model do not agree with those derived from the on-lattice model [\emph{cf.} Eq. (3.1)--(3.2) in Ref.~\onlinecite{Burger:2010gb}]. In particular, the lattice model does not include the negative terms in the diagonal entries of $\bf D$ (which are important to explain self-diffusion as we shall see below) and the advection term $\bf F$ is linear and does not include the difference ${\bf f}_b -  {\bf f}_r$. It would be interesting to study whether hopping rules can be given on a discrete lattice which produce the model \eqref{model_matrix} at a continuum level. 

\subsection{Numerical simulations} \label{sec:numericaltwo}

The particle-based description \eqref{sde2} of the problem, consisting of  $dN-$coupled stochastic differential equations (SDE), is used as a benchmark solution to test the validity of the reduced continuum model  \eqref{finalmodel_compact} for the marginal densities $b$ and $r$. To this end, we compare the solution of \eqref{finalmodel_compact} (obtained with the method of lines using a second-order finite-difference discretization for the spatial derivatives, in the spirit of the positivity-preserving finite-difference scheme proposed in Ref.~\onlinecite{Zhornitskaya:2000hr}), with Monte Carlo (MC) simulations of the $2N-$coupled SDE \eqref{sde2} in two dimensions ($d=2$). For the MC simulations, the reflective boundary conditions on $\partial \Omega$ are implemented as in Ref.~\onlinecite{Erban:2007we}, namely, the distance that a particle has travelled outside the domain is reflected back into the domain. The particle-particle overlaps are implemented similarly as follows. The difference $\epsilon - \| {\bf X}_i(t+ \Delta t) - {\bf X}_j(t + \Delta t) \|$ corresponds to the distance that particles have penetrated each other illegally. Then we suppose that each particle has travelled the same illegal distance, and we separate them accordingly along the line joining the two particles' centers. 
To test the importance of steric interactions, we also compare with the corresponding solutions with $\epsilon_b = \epsilon_r = 0$.

\begin{figure*}[htb]
\begin{center}
\includegraphics[height= 0.43\linewidth]{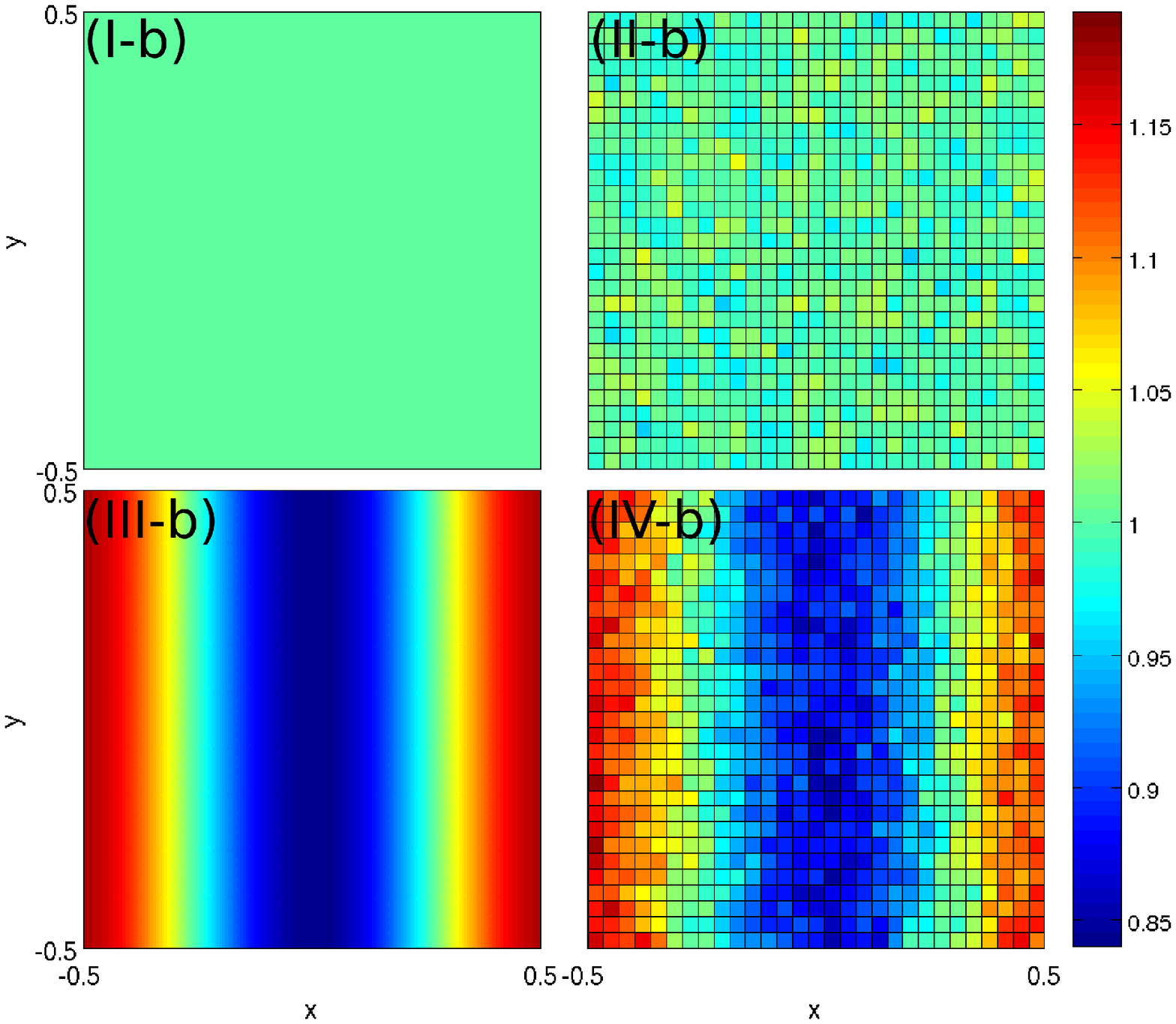} \hfill \includegraphics[height = 0.43\linewidth]{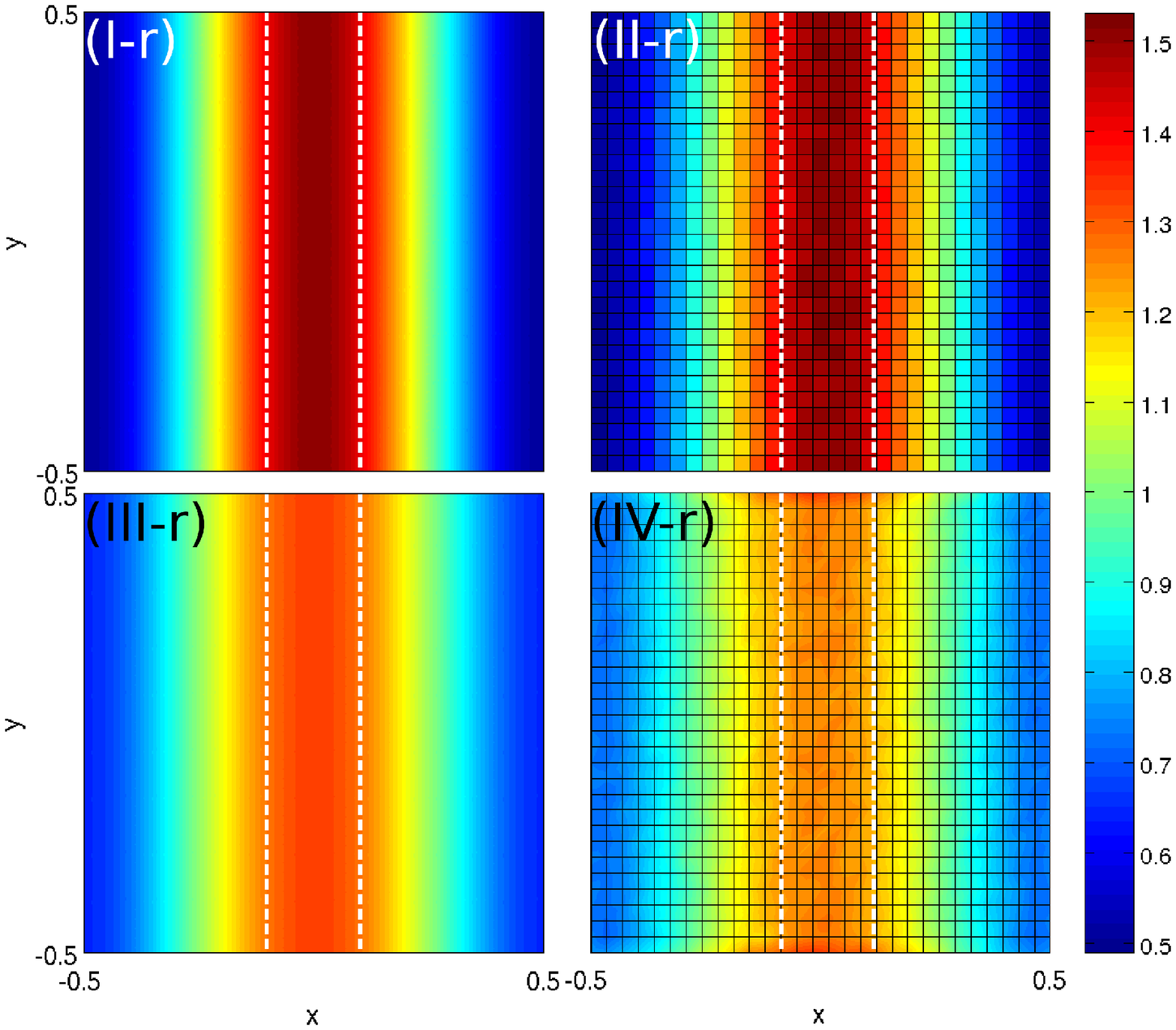}
\caption{Marginal densities $b({\bf x},t)$ ([I--IV]-b) and $r({\bf x},t)$ ([I--IV]-r) at time $t=0.05$ with initial data $b_0({\bf x}) = 1$ uniform and $r_0({\bf x})$ normally distributed in the $x$-axis,  ${\bf f}_b ={\bf f}_r = 0$, $D_b = D_r = 1$, $N_b=100$ and $N_r = 300$. 
(I) Solutions $b({\bf x},t)$ and $r({\bf x},t)$ of Eq. \eqref{2:point} for point particles ($\epsilon_b = \epsilon_r = 0$). (II) Histograms for $\epsilon_b = \epsilon_r = 0$. (III) Solutions $b({\bf x},t)$ and  $r({\bf x},t)$ of Eq. \eqref{finalmodel_compact} for finite-size particles ($\epsilon_b= 0.01, \, \epsilon_r = 0.02$). (IV) Histograms for $\epsilon_b = 0.01, \, \epsilon_r = 0.02$. Histograms computed from $10^4$ realizations of Eq. \eqref{sde2} with $\Delta t = 10^{-5}$. All four plots on the left and  right have respectively the same color bar.}
\label{fig:fig1}
\end{center}
\end{figure*}

Figure \ref{fig:fig1} shows the results of a time-dependent simulation with $ \bo {\mathrm f}_b(\bo {\mathrm x}) = \bo {\mathrm f}_r(\bo {\mathrm x}) \equiv 0$,  $\Omega = [-1/2,1/2]^2$, 
$N_b = 100$, $N_r = 300$,  $\epsilon_b = 0.01$, $\epsilon_r = 0.02$, and  $D_b =D_r= 1$. Initially, the blue particles are uniformly distributed, $b_0({\bf x}) = 1$, and the red particles are normally distributed in the $x_1$-axis with zero mean and standard deviation 0.1, $r_0({\bf x}) = f(x; 0, 0.1^2)$, where $f(x; \mu, \sigma^2)$ is the one-dimensional Gaussian (truncated and normalized so that its integral over $\Omega$ is one). The figures correspond to time $t=0.05$ and the simulation time step is set to  $\Delta t = 10^{-5}$. 

The theoretical predictions for both point and finite-size particles compare very well with their simulation counterparts, while steric effects are clearly appreciable even though the volume fraction of particles is only 0.102. The initial uniform density of the blue particles does not change in time when size-exclusion effects are ignored [\figref{fig:fig1} (I-b)], while it does diffuse towards the domain edges $x \pm 1/2$ when they are not [\figref{fig:fig1} (III-b)] due to the non-uniform density of red particles. More details on this effect will be given in \secref{sec:diffusions}. On the other hand, the red particles' initial profile, in which particles are concentrated in the central band, spreads faster when excluded-volume effects are included [\figref{fig:fig1} (III-r)] than when they are not [\figref{fig:fig1} (I-r)], indicating that the overall collective diffusion of the red species is enhanced.  

\section{Diffusion coefficients in a two-component system}
\label{sec:diffusions}
There exist several characterizations of diffusion in a system of finite-size particles. 
In the case of a single species, there is the collective diffusion coefficient, which
describes the evolution of the total concentration of the species, and the self-diffusion coefficient, which describes the evolution of a single tagged particle.\cite{Hanna:1982gi} For two or more species, a third coefficient, the cross-diffusion, expresses the motion of one species under a concentration gradient of the other species.\cite{Buzatu:2007fm}

\subsection{Collective diffusion}
The collective diffusion coefficients (also known as main or principal diffusion co\-ef\-fi\-cients \cite{Zielinski:2001tb}) are the diagonal entries of the diffusion matrix $\bf D$ in \eqref{diffusionmatrix}. For instance, the collective diffusion of the blue species is
\begin{equation}
\label{Dcollectiveb}
D_{bb} \bo(b ({\bf x},t), r ({\bf x},t) \bo) = D_{b} \left[ 1 + \epsilon_b^d (N_b-1) \alpha b({\bf x},t) - \epsilon_{br}^d N_r \gamma_b r({\bf x},t) \right].
\end{equation}
The first term is the diffusion of a free blue particle. The second term, which is proportional to the excluded-volume created by the blue particles, always enhances the collective diffusion. This enhancement is due to \emph{biasing} the random walk---in a gradient of $b$ you are more likely to move towards the low-density region. In contrast, the third term (related to the excluded-volume created by the red particles) reduces the collective diffusion of a blue particle. As expected, setting $N_r = 0$ in \eqref{Dcollectiveb} yields the collective diffusion coefficient for a single species of finite-size particles, see Eq. (13) of Ref.~\onlinecite{Bruna:2012cg}. Now, with the two species in play the collective diffusion coefficient displays a compromise between the enhancement due to the finite-size interactions within your own species and the diminishment due to the presence of the other species. 

\subsection{Cross-diffusion} \label{sec:crossdiff}
The cross-diffusion coefficients are the off-diagonal entries of the diffusion matrix  $\bf D$ in \eqref{diffusionmatrix}, which are always non-negative. For instance, the cross-diffusion coefficient of a blue particle across the red particles is
\begin{equation}
D_{br}\bo (b({\bf x},t) \bo ) = D_b \epsilon_{br}^d N_r \beta_b  b({\bf x},t).
\end{equation}
This term represents a drift on the blues density $b$ due to gradients in the reds density $r$.  We note that the name ``cross-diffusion'' to refer to such a term might be confusing, since it is really a drift, but this is a common terminology in the literature.\cite{Buzatu:2007fm,Hittmeir:2011bd}

\subsection{Self-diffusion}
The self-diffusion coefficient is different to the collective and cross-diffusion coefficients in that it is a diffusion coefficient intrinsically attached to an individual {\em tagged} particle, and which may be  related to its mean-square displacement. In contrast, the previous two coefficients relate a diffusive flux to the concentration gradient of many particles,\cite{Mazo:2002wk} that is, to the total concentration of $N_b$ or $N_r$ particles. Because the self-diffusion is a macroscopic property of an individual particle, its analysis in the current framework requires us to tag a single particle in the population-level model. We can do this by coloring one particle (the tagged particle) in red, $N_r=1$, leaving the remaining $N-1$ particles to be blue particles.

Setting $N_r=1$ and $N_b = N$, $D_b=D_r =1$, $\epsilon_b = \epsilon_r = \epsilon_{br} = \epsilon$, and ${\bf f}_i \equiv 0$ in \eqref{finalmodel_compact}, gives 
\begin{subequations}
\label{model_tagged}
\begin{align}
\label{tagged_b}
\frac{\partial {b}}{\partial t} ( {\bf x}, t)  &=  
\bo \nabla_{{\bf x}}  \cdot \left[  \bo \nabla_{{\bf x}} 
{ b}  + (N-1)  \epsilon^d \alpha \,    b  \bo \nabla_{{\bf x}} { b}   +   \epsilon^d \beta \, {b} \bo \nabla_{{\bf x}} {r} - \epsilon^d \gamma \, r \bo \nabla_{{\bf x}} {b}  \right], \\
\label{tagged_r}
\frac{\partial {r}}{\partial t} ( {\bf x}, t) &=    
\bo \nabla_{{\bf x}}  \cdot  \left [ \bo \nabla_{{\bf x}} 
{ r} - N \epsilon^d \gamma \, b \bo \nabla_{{\bf x}} r + N \epsilon^d  \beta \, r \bo \nabla_{{\bf x}} {b}    \right], 
\end{align}
where
\begin{equation}
\label{tagged_coef}
\alpha = \frac{2(d-1)\pi}{d},  \qquad   \beta =    \frac{(2d-1)\pi}{d},  \qquad 
\gamma =  \frac{\pi}{d},
\end{equation}
for $d=2$ or 3. 
\end{subequations}
Then the self-diffusion coefficient of the tagged red particle is
\begin{equation}
\label{self-diffusion_red}
D^s \bo ( b ( {\bf x}, t)   \bo ) = 1 - N \epsilon^d  \gamma b ( {\bf x}, t).
\end{equation}
Hence we find that the self-diffusion coefficient decreases for increasing excluded volume or, in other words, that it is reduced relative to point particles.

Let us compare this to the one-species collective diffusion coefficient.  
Since the red particle is identical to all the blue particles, we untag it so that if the initial densities are the same then $r \equiv b$ ( $=p$, say) and both equations \eqref{tagged_b} and \eqref{tagged_r} give
\begin{align}
\frac{\partial p}{\partial t} ( {\bf x}, t) =    
\bo \nabla_{{\bf x}}  \cdot  \left ( \bo \nabla_{{\bf x}} 
p + N \epsilon^d \alpha \, p \bo \nabla_{{\bf x}} p    \right) , 
\end{align}
since $\beta- \gamma = \alpha$. The diffusion coefficient of $p$ is then 
\begin{equation}
\label{collective_single}
D^c\bo ( p ( {\bf x}, t)   \bo )  = 1+ N \epsilon^d \alpha p ( {\bf x}, t), 
\end{equation}
which coincides with the effective collective diffusion coefficient in \eqref{fpN_one}.  Thus the collective diffusion coefficient $D^c$ is increased relative to point particles. This apparent contradiction may be understood as follows: the diffusion of any particular particle is impeded by its collisions with other particles. However, these collisions bias the random walk towards areas of low particle density, so that the overall spread of all particles is faster. 
When we look at the equation for the tagged red \eqref{tagged_r} the diffusion is \emph{reduced} relative to point particles, but there is also the drift term due to the gradient in the blues' density. The latter term is the dominant one when the blues and reds are the same species, since its coefficient is $(2d-1)$ times larger than the self-diffusion coefficient. The end result is that when the two are combined into a single term, the collective diffusion is {\em enhanced} relative to point particles, as seen in \eqref{collective_single}. 

Written in terms of the volume concentration $c = \phi p$, where $\phi = \pi N \epsilon ^d / 2d$ is the volume fraction, the self-diffusion coefficient $\eqref{self-diffusion_red}$ and collective diffusion coefficient \eqref{collective_single} read
\begin{align}
\label{diff_coeffcompare}
D^s(c) = 1- 2c, \hspace{1.5cm}
D^c(c)  = 1 + 4(d-1) c.
\end{align}
These expressions are in agreement with the values found in the literature using different methods.\cite{Ackerson:1982ti,Hanna:1982gi} Note that the self-diffusion coefficient is independent of the problem dimension unlike the collective diffusion coefficient (but note also that it is not defined for one-dimensional systems, because the hard-core interaction restricts the allowed motions in one dimension \cite{Ackerson:1982ti}). 

\subsection{Experimental measurements of diffusion coefficients}
\label{sec:experiment_measure}

\subsubsection{Mean squared displacement (MSD)}
The self-diffusion coefficient of a tagged particle can be described by using the particles' mean-square displacement (MSD), defined by MSD$(t) = \langle \|{\bf X}_i(t)  - {\bf X}_i(0)\|^2 \rangle$, where ${\bf X}_i(t)$ is the position of the $i$th particle at time $t$ and the angular bracket denotes an ensemble average (using that all particles are physically identical). If we keep the convention of  coloring in red the tagged particle, the following relation is obtained from the second moment of its PDE \eqref{tagged_r} (with zero drifts):
\begin{equation*}
(\partial  /\partial t) \, \text{MSD}(t) = 2d D^s(\phi),
\end{equation*}
where $d$ is the problem dimension, $D^s$ is given in \eqref{diff_coeffcompare}, and $c \equiv \phi$ for the system in equilibrium. This relation, known as the Einstein relation, is more commonly expressed as 
\begin{equation}
\label{msd_Ds}
D^s(\phi) = \lim_{t \to \infty}\text{MSD}(t)/(2dt).
\end{equation}
It relates a macroscopic quantity, the self-diffusion coefficient, with a microscopic quantity, the mean-square displacement. The latter can be measured in stochastic simulations of the particle system, which we shall use to check the validity of \eqref{diff_coeffcompare}. 

We perform Monte Carlo (MC) simulations of the $2N$-coupled SDE \eqref{sde2} in a two-di\-men\-sion\-al periodic box of 400 disks ($N=400$) under no external force (${\bf f}_b = {\bf f}_r \equiv 0$) and uniform initial distribution [$P_0(\vec x) =1$]. In order to achieve a better quantitative comparison, we employ the event-driven Brownian-dynamics (ED--BD) simulation scheme based on De Michele's algorithm.\cite{Scala:2007fd} We consider different volume fractions $\phi$, ranging between 0 and 0.1, and the particles' size $\epsilon$ is chosen to achieve the desired volume fraction.  \figref{fig:fig2}(a) shows an example plot of the mean square displacement of the disks as a function of time when $\phi=0.01$. Note that MSD$(t)$ varies linearly with $t$ for all times, indicating that the system is already in the stationary at the initial simulation time. This is because we have thrown away the thermalization transition of our simulations. From its slope at long times the self-diffusion coefficient may be extracted using  \eqref{msd_Ds}. Varying the volume fraction in the simulation, we obtain points $\bo (\phi, D^s(\phi) \bo )$ which are plotted in \figref{fig:fig2}(b) as red circles. The theoretical curve $D^s(\phi)$, shown as a black dashed-line, compares well with the measured values.

\def \scc {0.6}
\def \scl {1}
\begin{figure}[htb]
\unitlength=1cm
\begin{center}
\psfragscanon
\psfrag{a}[][][\scl]{$(a)$} \psfrag{b}[][][\scl]{$(b)$} 
\psfrag{t}[][][\scl]{$t$} \psfrag{MSD(t)}[][][\scl]{MSD$(t)$}
\psfrag{f}[][][\scl]{$\phi$} \psfrag{D(f)}[][][\scl]{$D^s(\phi)$}
\includegraphics[height=0.34\textwidth]{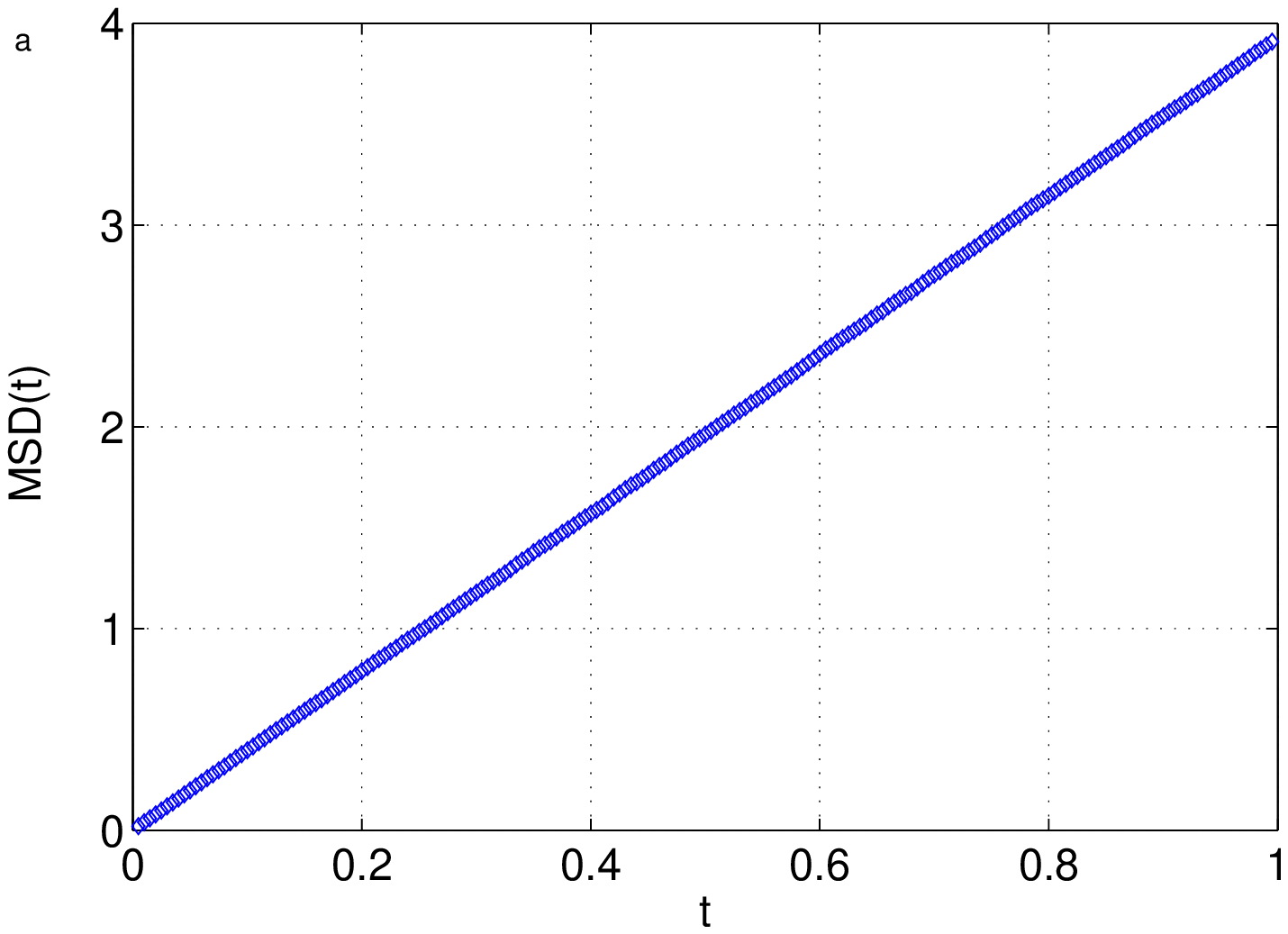} \hfill \includegraphics[height=0.34\textwidth]{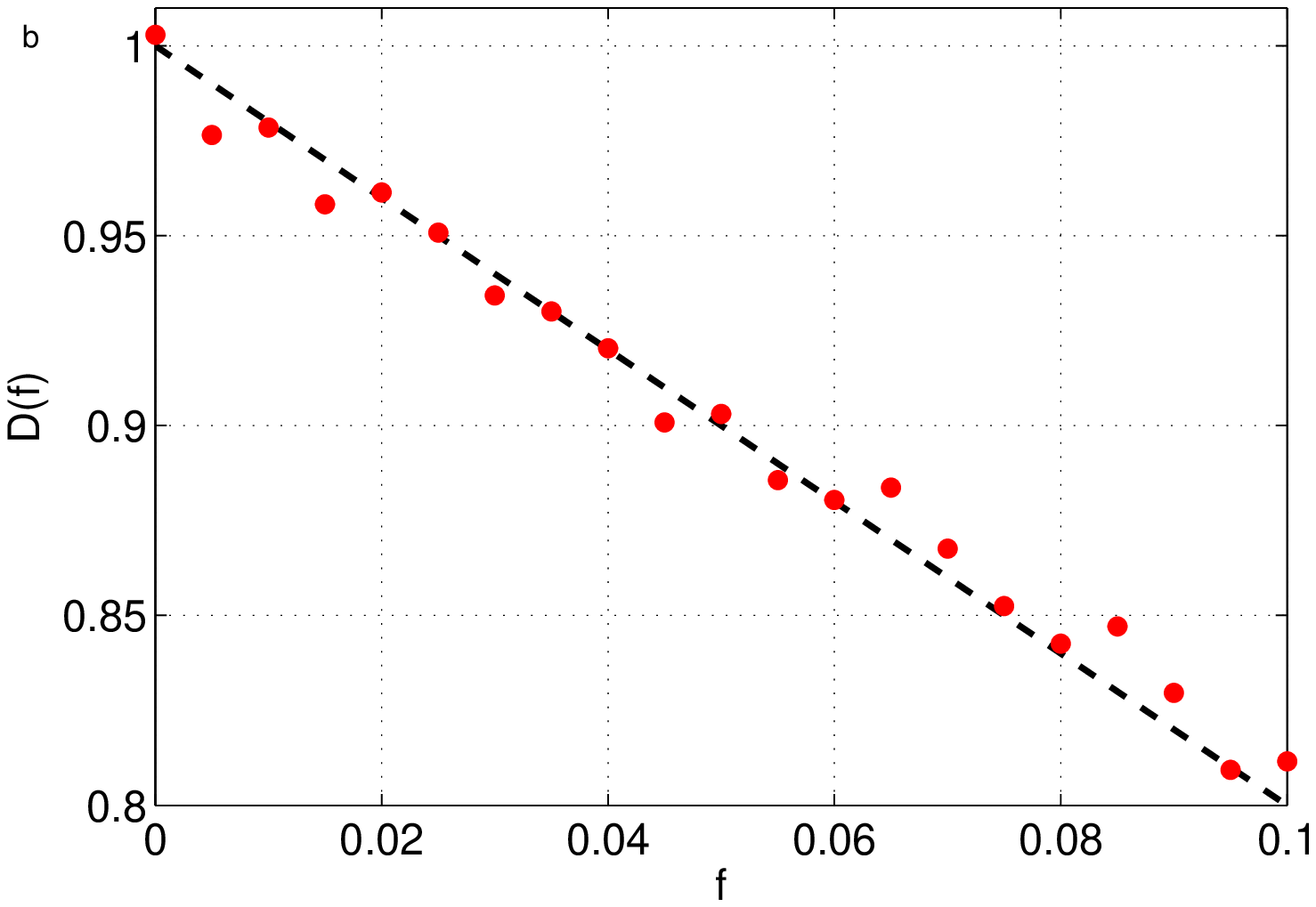}
\caption{Results from stochastic simulation of a two-dimensional periodic system with $N=400$ hard disks and variable volume fraction $\phi$ (achieved by changing the particles' diameter $\epsilon$). $(a)$ Mean-square displacement MSD$(t)$ for a volume fraction $\phi=0.01$. (b) Self-diffusion coefficient $D^s(\phi)$ for volume fractions of up to 10\%. Measured values from stochastic simulations using \eqref{msd_Ds} (red circles) and theoretical prediction \eqref{diff_coeffcompare} (dashed line). }
\label{fig:fig2}
\end{center}
\end{figure}

\subsubsection{Fluorescence recovery after photobleaching (FRAP)}
\label{sec:frap}

Fluorescence recovery after photobleaching (FRAP) is an experimental technique for measuring the mobility of fluorescent particles. Since the introduction of noninvasive fluorescent tagging with fluorescent proteins, this technique has become widely used to study protein dynamics in living cells.\cite{Carrero:2003ea} In a FRAP experiment, a subregion of the cell (the sample volume) is photobleached with a laser beam, causing the molecules contained in it to loose their fluorescence irreversibly. As a result, two species are formed, the photobleached molecules (inside the sample volume) and the fluorescent molecules (outside the sample volume). Subsequently, the recovery of the equilibrium from this perturbed initial state is monitored by measuring the fluorescence intensity in the sample volume. The resulting curve of intensity against time, which can be related to the concentration of the fluorescent species, is then used to estimate the overall mobility of the molecule.\cite{Braga:2007gt}

Several models to fit simulated curves to experimental data have been proposed, most of which assume the transport mechanism of proteins to be diffusive.\cite{Carrero:2003ea} The standard method is based on the work of Axelrod \emph{et. al.}\cite{Axelrod:1976ts}, in which a linear diffusion equation is used to model the fluorescence recovery in a two-dimensional infinite domain. Since the motion of molecules inside the cell can be influenced by many complex interactions such as excluded-volume effects or binding events, the estimated diffusion coefficient is often referred to as the effective or apparent diffusion coefficient.\cite{Dix:2008gy,Braga:2007gt,Carrero:2003ea} Modifications of the diffusion model of Axelrod to estimate parameters other than the diffusion coefficient, such as the immobile fractions or binding rates,  or to account for multiple diffusive species, have also been used.\cite{GonzalezPerez:2011ty,Braga:2007gt}

In contrast with the MSD, which is a single-particle tracking method, FRAP is an ensemble-averaged method describing the averaged diffusive properties of many fluorescent particles.\cite{Dix:2008gy} As a result, even in the simplest scenario of diffusion alone (without binding or immobile particles), the diffusion measurements from MSD or FRAP will differ unless particles are interaction-free (since one is measuring self-diffusion and the other some for of collective diffusion). 
Nevertheless, FRAP experiments have beed extensively used to measure the diffusion coefficient of \emph{one} molecule, and erroneously seen as an alternative to the MSD  (for example, when the latter is not feasible due to limitations in labeling or rapid diffusion).\cite{Dix:2008gy,Saunders:2012jl} 

Therefore, care should be taken to interpret the diffusion coefficient estimated from FRAP experiments. The first point to note is that the FRAP diffusion coefficient cannot be identified as the self-diffusion coefficient of the molecule, often termed the \emph{anomalous} subdiffusion.\cite{Dix:2008gy} Instead, it characterizes the mobility of the whole fluorescent population. Secondly, the standard linear diffusion equation ignores interactions between the photobleached and fluorescent species. Is not clear whether a ``pure'' diffusion measure can be obtained from the FRAP experiments or, as we have seen in \secref{sec:crossdiff}, it is a combination of collective diffusion and drift due to gradients of the photobleached species (cross-diffusion) that it is in fact measured.

The two-species model \eqref{finalmodel_compact} can be used to model the FRAP experiment. For example, consider the simple setting a two-dimensional unit square domain, pure diffusion and a circular photobleached area (sample area) of radius $w$,\cite{Axelrod:1976ts} with $\epsilon\ll w \ll 1$. The system is initially in equilibrium, so that the concentration is uniform in $\Omega$ and equal to the volume fraction $\phi = \pi N \epsilon^d/2d$. A laser of intensity $I$ causes the portion of particles contained in $\|{\bf x}\| <w$ to be photobleached, which we identify as the blue particles. The fluorescent particles comprise the red species. We consider volume concentrations rather than probability densities to relate them to FRAP measurements. The number of bleached particles is $N_b = \pi w^2 N$ and their initial volume concentration is $c_b({\bf x},0) = \phi$ for $\|{\bf x}\| <w$ and 0 otherwise. For the fluorescent particles, $N_r = N(1-\pi w^2)$ and $c_r({\bf x},0) = \phi$ for $\|{\bf x}\| >w$. Since initially all particles are identical, they have equal diffusion coefficient $D$. Using that $N_r$ is large such that $N_r-1 \approx N_r$, the equation for the fluorescent (red) species \eqref{finalmodelc_r} reads
\begin{align}
\label{eq_frap}
\frac{\partial c_r}{\partial t} ({\bf x},t) = D \bo \nabla_{\bf x} \cdot \left\{ [1 + 4 (d-1)c_r -2c_b] \bo \nabla_{\bf x} c_r + 2(2d-1) c_r \bo \nabla_{\bf x} c_b \right\},
\end{align}
where $c_r = \pi N_r \epsilon^d r/2d$. An analogous equation is obtained for the bleached (blue) species. Solving this system of equations for $c_r$ and $c_b$ with the initial conditions described above until equilibrium is reached [uniform concentrations $c_r({\bf x},\infty) = (1-\pi w^2)\phi$ and $c_b ({\bf x},\infty) = \pi w^2 \phi$], the theoretical recovery curve, which is related to the integral over the sample area of $I({\bf x}) c_r({\bf x}, t)$,\cite{Carrero:2003ea} could be compared with the experimental recovery curve.

\section{Basic Properties}
\label{sec:basicprops}

In this section we discuss some basic properties of the cross-diffusion model \eqref{finalmodel_compact}, such as its free-energy functional, equilibria and stability of solutions. We restrict ourselves to the case when the force fields are potential forces, that is, ${\bf f}_b  ( {\bf x}) = - \bo \nabla_ { {\bf x}} V_b ( {\bf x}) $ and $ {\bf f}_r  ( {\bf x}) = - \bo \nabla_ { {\bf x}} V_r ( {\bf x})$; \emph{i.e.}, we consider the system
\begin{subequations}
\label{finalmodel_compactV}
\begin{align}
\label{finalmodelcV_b}
\begin{aligned}
\frac{\partial {b}}{\partial t} ( {\bf x}, t)  =  
\bo \nabla_{{\bf x}}  \cdot \Big( &  D_b \left[ 1 +
(N_b-1)  \epsilon_b^d \alpha    b \right] \bo \nabla_{{\bf x}} { b} + \bo \nabla_{ \bf x} V_b( {\bf x})  b     \\ 
& +  N_r \epsilon_{br}^d \big \{ D_b \left[ \beta_b \, {b} \bo \nabla_{{\bf x}} {r} - \gamma_b {r}\bo \nabla_{{\bf x}} {b} \right] + \gamma_b \bo \nabla_{ \bf x} \left[  V_r( {\bf x}) -  V_b( {\bf x}) \right] br\big \} \Big),   
\end{aligned}
\\
\label{finalmodelcV_r}
\begin{aligned}
\frac{\partial {r}}{\partial t} ( {\bf x}, t) =    
\bo \nabla_{{\bf x}}  \cdot  \Big(& D_r \left[ 1 + (N_r-1)  \epsilon_r^d \alpha   r  \right] \bo \nabla_{{\bf x}} 
{ r} +   \bo \nabla_{ \bf x} V_r( {\bf x})   r   \\
&+ N_b \epsilon_{br}^d  \big \{ D_r \left[ \beta_r \, {r} \bo \nabla_{{\bf x}} {b} - \gamma_r b\bo \nabla_{{\bf x}} r \right] + \gamma_r \bo \nabla_{ \bf x} \left[  V_b( {\bf x}) -  V_r( {\bf x}) \right]  br \big \} \Big), 
\end{aligned}
\end{align}
in $\Omega\times \mathbb R^+$ with no-flux boundary conditions on $\partial \Omega\times \mathbb R^+$ and initial conditions
\begin{equation}
\label{finite_initial}
b({\bf  x},0) = b_0({\bf  x}), \qquad r({\bf  x},0) = r_0({\bf  x}) \qquad \text{in} \qquad \Omega.
\end{equation}
\end{subequations}
The coefficients $\alpha$, $\beta_i$ and $\gamma_i$ ($i= b, r$) are all non-negative and given in \eqref{coef_23d}.

\subsection{Gradient flow structure and free energy}
\label{sec:grad flow-entropy}

In Ref.~\onlinecite{Bruna:2012cg} we found that, when ${\bf f}({\bf x})=-\nabla_{{\bf x}} V({\bf x})$, the equation for one species \eqref{fpN_one} can be written as gradient flow \cite{Burger:2010gb}
\begin{subequations}
\label{gradflow}
\begin{equation}
\label{gradflow1}
\frac{\partial p}{ \partial t} + \bo \nabla_{\bf x} \cdot (p {\bf u}) = 0,
\end{equation}
with ${\bf u} = - \nabla_{{\bf x}}  [ D \log p +  2D \alpha_d (N-1) \epsilon ^d p  +  V({\bf x})]$. (In Ref.~\onlinecite{Bruna:2012cg}, $\bf u$ and $\mathcal F(p)$ were defined as $-\bf u$ and $F(p)$ respectively.) Here  $\bf u$ can be thought of as a ``flow'' down the gradient of the free energy $\mathcal F(p)$ associated to equation \eqref{fpN_one}, ${\bf u} = - \bo \nabla_{\bf x} \frac{\delta \mathcal F}{\delta p}$, with
\begin{equation}
\label{entropy1}
\mathcal F(p) =  \int_{\Omega} D \left[ p \log p +  \alpha (N-1) \epsilon ^d p^2 \right] \ud {\bf x} +  \int_{\Omega} V({\bf x}) p \, \ud {\bf x}.
\end{equation}
\end{subequations}
The first integral corresponds to the internal energy, which increases with excluded-volume effects, and the second integral is the potential energy. Then the free energy is non increasing in time when evaluated along a solution of \eqref{fpN_one}.\cite{Carrillo:2001ee}
The gradient flow structure \eqref{gradflow} is very useful since it brings more tools to study the trend to equilibrium,\cite{Villani:2002wj} with the free energy functional \eqref{entropy1} ``encoding'' all the properties of the flow. 

The scalar gradient flow structure \eqref{gradflow1} becomes, in the case of two species, \cite{Burger:2010gb}
\begin{align}
\label{grad_flow}
\frac{\partial}{\partial t}
\begin{pmatrix} b \\ r \end{pmatrix} ({\bf x},t)  = 
\bo \nabla_{\bf x} \cdot \left[  {\bf M} \bo \nabla_{\bf x} 
\begin{pmatrix} \partial _{ b} \mathcal F  \\ \partial _{r} \mathcal F \end{pmatrix} \right],
\end{align}
where $\mathcal F = \mathcal F(b,r)$ is, again, a scalar free-energy functional and ${\bf M} = {\bf M} ( b,  r)$ is a two-dimensional matrix denoted the mobility matrix (which must be positive semi-definite from the definition of free energy). In this section we examine under which conditions the two-species system \eqref{finalmodel_compactV} admits an explicit gradient flow representation \eqref{grad_flow} valid to $\mathcal O(\epsilon_b^d, \epsilon_r^d)$, bearing in mind that $\epsilon_b, \epsilon_r, \epsilon_{br}\ll 1$. It should be pointed out that the transformation of \eqref{finalmodelcV_b}--\eqref{finalmodelcV_r} into the structure \eqref{grad_flow} is not straightforward in general. In the following we present two situations in which our system admits a gradient flow structure: for a large number of particles (\subsecref{sec:largenum}) and  when the drift terms become zero (\subsecref{sec:nodrift}).

\subsubsection{Large number of particles approximation}
\label{sec:largenum}

Assume that the number of blue and red particles is large such that $N_b -1 \approx N_b$ and $N_r -1  \approx N_r$, and that the two species have the same diffusion coefficient (which we can set to unity without loss of generality). Then the system \eqref{finalmodel_compactV} for the densities $b$ and $r$ can be rewritten in terms of the number densities $\hat b = N_b b  $ and $\hat r = N_r  r $ in the gradient flow form \eqref{grad_flow}, with free-energy functional
\begin{subequations}
\label{tote_ex1}
\begin{align}
\label{entropy_ex1}
\mathcal F (\hat b, \hat r) \! = \int_{\Omega}  \hat b \log \hat b +  \hat r \log \hat r + \hat b V_b + \hat r V_r +\frac{\alpha}{2} \left( \epsilon_b^d \, \hat b^2 + 2 \epsilon_{br}^d \, \hat b\hat r  + \epsilon_r^d \, \hat r^2 \right) \ud  {\bf x},
\end{align}
and mobility matrix
\begin{align}
\label{mobility_ex1}
\renewcommand{\arraystretch}{1.0}
{\bf M} (\hat b, \hat r) = 
\begin{pmatrix}  \hat b (1 - \gamma \epsilon_{br}^d \hat r) & \gamma \epsilon_{br}^d \hat b \hat r  \\
\gamma \epsilon_{br}^d \hat b \hat r &      \hat r (1 -  \gamma \epsilon_{br}^d \hat b)  \end{pmatrix},
\end{align}
where $\alpha = 2 (d-1) \pi/d$ and $\gamma = \pi/d$ is the simplified version of  $\gamma_i$ ($i = b, r$) in \eqref{coef_23d} when the diffusivities of blues and reds are equal. [The coefficient $\beta_i$  disappears in Eq. \eqref{tote_ex1} using that $\beta_i \equiv \alpha + \gamma$ when $D_b = D_r = 1$.] 
\end{subequations}
The mobility matrix $\bf M$ is positive definite if and only if
\begin{equation}
\label{upperbound_ex1}
\gamma \epsilon_{br}^d (\hat b + \hat r) < 1,
\end{equation}
using that $\hat b, \hat r \ge 0$. This upper bound on the total number density $\hat b + \hat r$ gives a limit of validity of the model, and must be satisfied pointwise in $\Omega$. Nevertheless, to get an approximate idea of the upper bound on volume fraction $\phi$, assume that the concentrations are constant, $\hat b = N_b$ and $\hat r = N_r$ and that $\epsilon_b = \epsilon_r $. Then, using that the volume fraction is $\phi = \frac{\pi}{2d} \epsilon_{br}^d (N_b + N_r)$, Eq. \eqref{upperbound_ex1} becomes $\phi<0.5$. 
Therefore, we see that our model for two species in the case of large number of particles and equal diffusivities would break down when the volume fraction reaches one half. 

\subsubsection{Zero potential}
\label{sec:nodrift}
Now consider the system \eqref{finalmodel_compactV} with zero potential, $V_b=V_r \equiv 0$. In this case our problem has also a gradient flow structure \eqref{grad_flow} with
\begin{subequations}
\label{finalmodel_compact_ex2}
\begin{align}
\label{finalmodel_compact_ex2_entropy}
\mathcal F (b, r) &= \int_{\Omega} \left[ b \log  b +   r \log r  + \frac{\alpha}{2}(N_b-1) \epsilon_b^d \,  \alpha b^2 +  \frac{\alpha}{2}(N_r-1) \epsilon_r^d \, r^2 + \Theta \,  \epsilon_{br}^d  b r \right] \ud  {\bf x},
\end{align}
and
\begin{align}
\label{finalmodel_compact_mobmatrix}
\renewcommand{\arraystretch}{1.2}
{\bf M} ( b,  r) =
\begin{pmatrix}
D_b b (1 - N_r\epsilon_{br}^d \gamma_b \, r )  \quad & D_b\epsilon_{br}^d (N_r\beta_b - \Theta)  b  r  \\ 
D_r \epsilon_{br}^d (N_b\beta_r - \Theta ) b  r \quad &       D_r r (1 - N_b\epsilon_{br}^d\, \gamma_r  b ) 
\end{pmatrix},
\end{align}
\end{subequations}
where $\Theta$ is a free parameter. There are two relevant cases for the gradient flow structure \eqref{finalmodel_compact_ex2} depending on the value given to $\Theta$. First, the free energy \eqref{finalmodel_compact_ex2_entropy} can be chosen to be the sum of the two one-species entropies like \eqref{entropy1} by setting $\Theta=0$. 
Second, the mobility tensor can be adjusted to be symmetric setting $\Theta = (N_b D_r\beta_r-N_r D_b\beta_b) / (D_r - D_b)$ provided $D_b\ne D_r$. (In case $D_b = D_r$, we can still obtain a symmetric mobility matrix while at the same time setting $\Theta = 0$ by rewriting \eqref{finalmodel_compact_ex2} in terms of the number densities $\hat b$ and $\hat r$ as in the previous subsection.)
The determinant of the mobility matrix \eqref{finalmodel_compact_mobmatrix} is
\[
\det {\bf M} ( b,  r)  = D_b D_r b r\left (  1-   \frac{2\pi}{d} \frac{(D_r N_b b + D_b N_r r)}{D_b + D_r} \epsilon_{br}^d  \right ) + \mathcal O(\epsilon_{br}^{2d}),
\]
where note that the free parameter $\Theta$ only enters at $\mathcal O(\epsilon_{br}^{2d})$. The mobility matrix is positive  definite if (neglecting the $\mathcal O(\epsilon_{br}^{2d})$ terms)
\begin{equation}
\label{upperbound_ex2}
\frac{2\pi}{d} \frac{(D_r N_b b + D_b N_r r)}{D_b + D_r}  \epsilon_{br}^d  < 1,
\end{equation}
using that $b, r> 0$. As before, if we assume that $b=r=1$ (we have reached the equilibrium), $D_b=D_r $ and $\epsilon_b = \epsilon_r$, the upper bound \eqref{upperbound_ex2} becomes $\phi<0.5$ as in the previous subsection. The point at which the mobility matrix becomes singular can be related with the stability of the equilibrium states of the system, see \secref{sec:linearstab}.

\subsection{Equilibria}
\label{sec:equilibria}
We compute the stationary solutions of the nonlinear diffusion model \eqref{finalmodel_compactV}, which we denote by $b_s$ and $r_s$. Note that, in the case of point particles ($\epsilon_b = \epsilon_r = 0$), the equilibria are trivial as the system for $b_s$ and $r_s$ decouples and we find
\begin{equation}
b_s ({\bf x}) = C_b \exp  [-V_b ({\bf x}) / D_b], \qquad r_s ({\bf x}) = C_r \exp [-V_r ({\bf x}) / D_r ],
\end{equation}
where $C_b$ and $C_r$ are the normalization constants, \emph{i.e.}, $C_i= \int_\Omega \exp [{-V_i ({\bf x}) / D_i}]  \, \ud {\bf x}$.
For finite-size particles, we first consider the cases for which we have found a gradient flow structure \eqref{grad_flow} in \secref{sec:grad flow-entropy}, and then we consider the general case in two ways: solving the steady states of \eqref{finalmodel_compactV} with a finite difference numerical scheme, and similarly the stationary distribution of the particle description \eqref{sde2} with stochastic simulations using the Metropolis--Hastings (MH) algorithm.\cite{Chib:1995ud}

\subsubsection{From the free-energy description}

One big advantage of the system with a gradient flow structure is that the steady states can be found in a straightforward manner by minimizing the free energy $\mathcal F$. If the mobility matrix ${\bf M}(b_s, r_s)$ is positive definite, then finding the steady states of \eqref{grad_flow} is equivalent to finding constants $\chi_b$ and $\chi_r$ such that
\begin{align}
\label{minim}
\partial_{b} \mathcal F(b_s, r_s)  = \chi_b, \hspace{1.5cm}
\partial_{r} \mathcal F(b_s, r_s)  = \chi_r, 
\end{align}
with $\int b_s = \int r_s = 1$. Note that the no-flux boundary conditions on $\partial \Omega$ are automatically satisfied by imposing $\partial_b \mathcal F$ and $\partial_r \mathcal F$ to be constant. 

In the case of no external potentials (\secref{sec:nodrift}), we see that the obvious constant states $b_s = r_s = 1$ are the only stationary solutions of \eqref{finalmodel_compact_ex2}. 
In the case of a large number of particles (\secref{sec:largenum}), the stationary solutions $\hat b_s$ and $\hat r_s$ (recall we used number densities for this case) are found by imposing \eqref{minim} on the associated free energy \eqref{entropy_ex1}. Then the problem reduces to
\begin{equation}
\label{ss_ex1}
\begin{aligned}
\log \hat b  + V_b + \alpha(\epsilon_b^d \hat b + \epsilon_{br}^d \hat r ) &= \chi_b,\\
\log \hat r  + V_r + \alpha( \epsilon_r^d \hat r + \epsilon_{br}^d \hat b ) &=  \chi_r, 
\end{aligned}
\end{equation}
such that $\int_\Omega \hat b_s = N_b$ and $\int_\Omega \hat r_s = N_r$. This problem can be solved with the Newton--Raphson method, discretizing $\Omega$ into $J$ grid points, approximating the normalization integrals with a quadrature, and solving for $2J+2$ unknowns ($\chi_b$, $\chi_r$, $\hat b_{k}$, and $\hat r_{k}$, for $i=1,\dots, J$).

\subsubsection{From the particle system via the Metropolis--Hastings algorithm} 

As we did with the time-dependent case, we now compare the stationary solution of the reduced macroscopic model \eqref{finalmodel_compactV} to the stationary distribution of the full microscopic model \eqref{twofp_eq}.
Under external potential forces, \eqref{fp_junts} becomes, on rearranging,
\begin{align}
\label{fp_full_potentials}
\frac{\partial P}{\partial t}(\vec x,t)  &= \sum_{i=1}^N D_i \bo \nabla_{{\bf x}_i} \cdot \left \{ P \, \bo \nabla_{{\bf x}_i} \! \left[\log P + V_i  ({\bf x}_i) / D_i \right] \right\},
\end{align}
where  $D_i = D_b$, $V_i= V_b$ for  $i \le N_b$ and $D_i = D_r$, $V_i = V_r$ otherwise. Denoting $\mathbb D$ the $N-$diagonal matrix with diagonal entries $D_i$, we can write
\begin{align} 
\label{fp_full_potentials2}
\frac{\partial P}{\partial t} + \vec  \nabla_{\vec x} ^\top \,\mathbb D\, \left (P \vec u \right ) = 0,
\end{align}
where $\vec u = - \vec \nabla_{\vec x}  [\log P + \sum_{i=1}^N V_i  ({\bf x}_i) /D_i ]$ is the ``flow velocity'' vector field. Using that $\mathbb D$ is non-singular and that no-flux boundary conditions hold, the stationary solution $P_s$  of \eqref{fp_full_potentials2} is $P_s(\vec x) = C\exp[-\mathcal H(\vec x)]$, where $\mathcal H(\vec x)$ is given by:
\begin{equation}
\label{2:metropolis1}
\mathcal H(\vec x) = \left\{ 
\begin{array}{l l}
\sum_{i=1}^N V_i  ({\bf x}_i) /D_i, \qquad  & \vec x \in \Omega_\epsilon^N,\\
\infty,& \text{otherwise}.
\end{array}
\right.
\end{equation}

\subsubsection{Examples}
\begin{figure*}[htb!]
\begin{center}
\includegraphics[height= 0.42\linewidth]{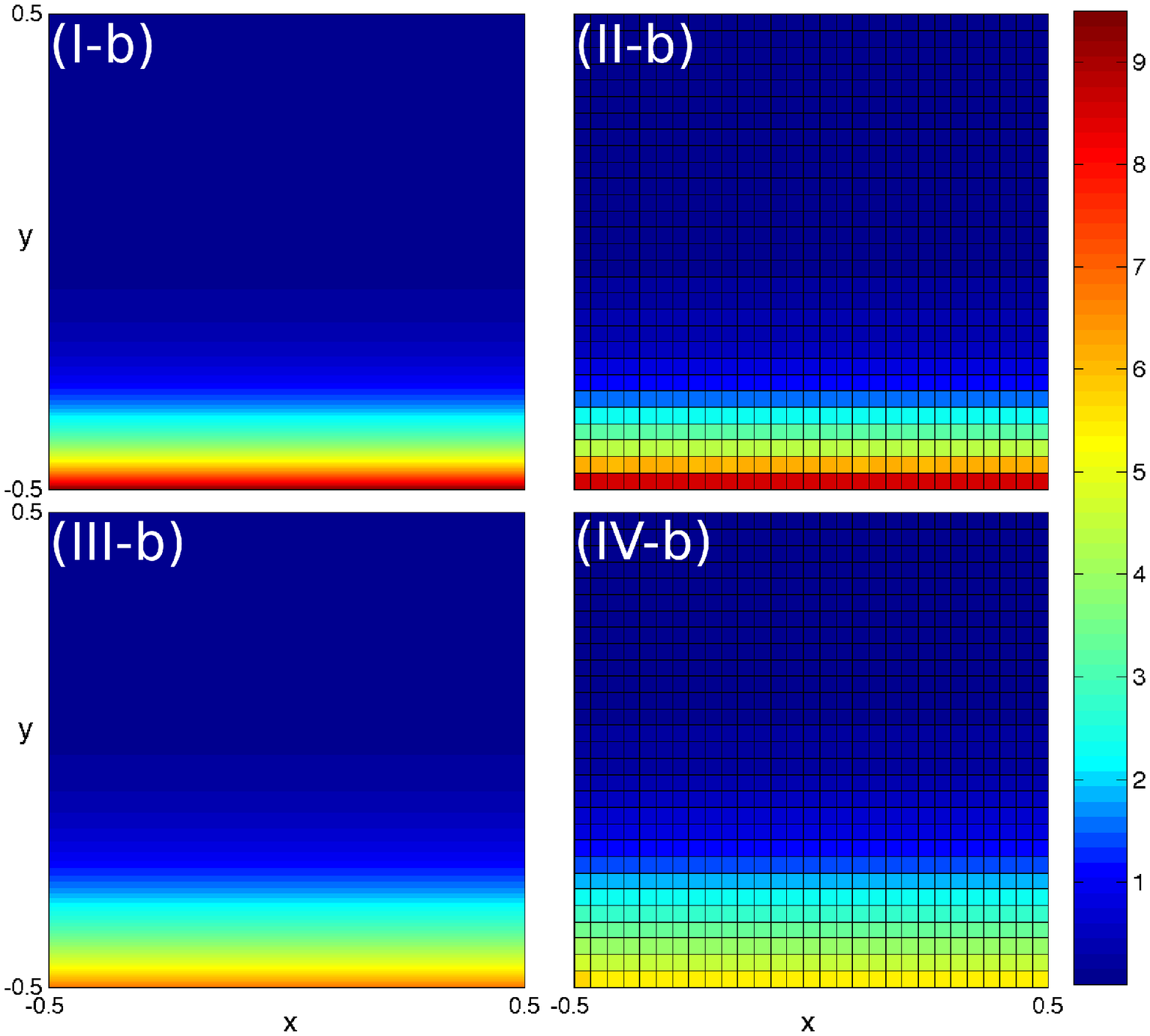} \quad \includegraphics[height = 0.42\linewidth]{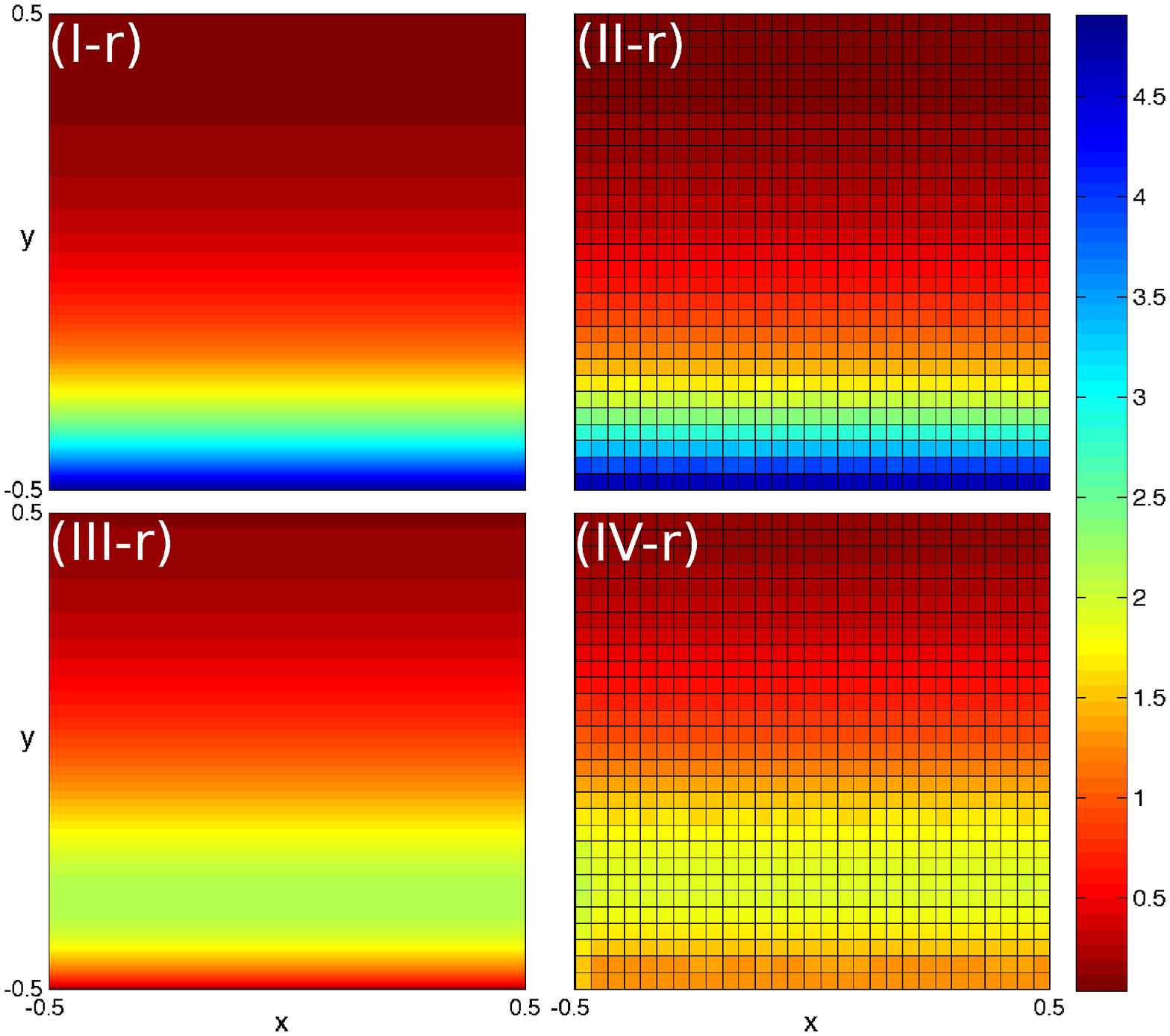} 
\caption{Stationary marginal densities $b_s({\bf x})$ ([I--IV]-b) and $r_s({\bf x})$ ([I--IV]-r) for point and finite-size particles, with $V_b = 10y$, $V_r = 5y$, $D_b = D_r = 1$, $N_b=600$ and $N_r = 200$.
(I) Solutions  $b_s({\bf x})$ and  $r_s({\bf x})$ of \eqref{ss_ex1} for point particles ($\epsilon_b = \epsilon_r = 0$). (II) Histograms for $\epsilon_b = \epsilon_r = 0$. (III) Solutions  $b_s({\bf x})$ and  $r_s({\bf x})$ of \eqref{ss_ex1}  for finite-size particles ($\epsilon_b = 0.01$,  $\epsilon_r = 0.015$). (IV) Histograms for $\epsilon_b = 0.01$,  $\epsilon_r = 0.015$. Histograms computed from $10^7$ steps of the MH algorithm. All four plots on the left and right have respectively the same color bar (note inverted color scale for reds and blues).}
\label{fig:fig3}
\end{center}
\end{figure*}

Two examples of stationary densities are shown in Figures \ref{fig:fig3} and \ref{fig:fig5}. The stationary solution of Eq. \eqref{finalmodel_compactV}, computed using \eqref{ss_ex1}, is compared with stochastic simulations of the stationary distribution of the  full $N$-particle SDE \eqref{sde2} with the MH algorithm.

\figref{fig:fig3} corresponds to the stationary solution under a `gravitational' force in the direction $-{\bf e}_y$, with $\Omega = [ -1/2, -1/2]$, $\epsilon_b = 0.01$, $\epsilon_r = 0.015$,  $D_b = D_r = 1$, $N_b=600$, and $N_r = 200$. We suppose that the blue particles (four plots on the left) are heavier than the red particles (four  plots on the right), and that therefore they feel a stronger force, ${\bf f}_b = -10 {\bf e}_y$ versus ${\bf f}_r = -5 {\bf e}_y$. While both blue and red particles accumulate at the bottom when finite-size effects are ignored [\figref{fig:fig3}(I-b) and \figref{fig:fig3}(I-r)], the blue particles accumulate at the bottom  [\figref{fig:fig3}(III-b)] and force the red particles upwards [\figref{fig:fig3}(III-r)] when they are not (note there is zero probability of finding a red particle at $y=-0.5$). The averages of $b_s$ and $r_s$ across $x$ are shown in \figref{fig:fig4}. The agreement between the model \eqref{ss_ex1} and the stochastic simulations is good in all cases, except in the region  near $y=-0.5$ for the red finite-size particles [compare \figref{fig:fig3}(III-r) and \figref{fig:fig3}(IV-r)]. A possible explanation for this disagreement is that the variability of $r_s$ near that boundary occurs in a region of size equal to the size of the histogram bins.  
\def \scc {0.8}
\def \scl {1}
\begin{figure}[htb!]
\unitlength=1cm
\begin{center}
\psfragscanon
\psfrag{V}[][][\scl]{} 
\psfrag{y}[][][\scl]{$y$} \psfrag{avx}[][][\scl]{$\langle p \rangle_x$} 
\psfrag{datadatadata1}[][][\scc]{$b_s$ point} 
\psfrag{datadatadata2}[][][\scc]{(sim.)} 
\psfrag{datadatadata3}[][][\scc]{$r_s$ point} 
\psfrag{datadatadata4}[][][\scc]{(sim.)} 
\psfrag{datadatadata5}[][][\scc]{$b_s$ finite size} 
\psfrag{datadatadata6}[][][\scc]{(sim.)} 
\psfrag{datadatadata7}[][][\scc]{$r_s$ finite size} 
\psfrag{datadatadata8}[][][\scc]{(sim.)} 
\includegraphics[height = 0.42\linewidth]{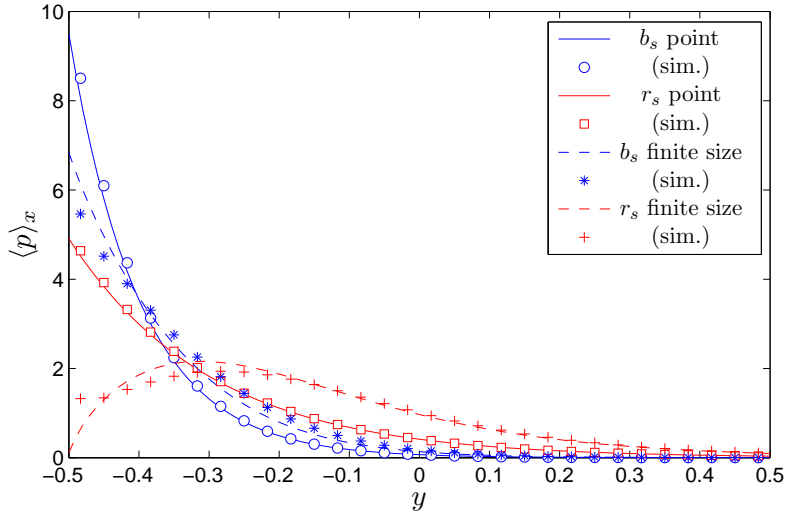}
\caption{Averaged stationary densities across $x$, $\langle b_s \rangle_x$ and $\langle r_s \rangle_x$, corresponding to the 8 cases shown in \figref{fig:fig3}. Comparison between stationary solutions of \eqref{ss_ex1} (curves) and histograms obtained from MH simulations (data points).}
\label{fig:fig4}
\end{center}
\end{figure}

\figref{fig:fig5} corresponds to the stationary solution under a symmetric bivariate Gaussian potential of the form
\[
\mathcal G({\bf x}; \mu, \sigma^2) = 1/(2\pi \sigma^2) \exp \{ -[(x - \mu)^2 +  (y - \mu)^2]/2\sigma^2\}.
\]
The parameters are $V_b = -0.1 \mathcal G({\bf x};0, 0.05)$, $V_r = 0.5 \mathcal G({\bf x};0.2, 0.05)$, $\epsilon_b =\epsilon_r = 0.02$, $D_b = D_r = 1$, $N_b= N_r = 400$ and $\Omega = [ -1/2, 1/2]$. For point particles, the stationary solutions preserve the radial shape and centers of their respective potentials $V_b$ and $V_r$, \emph{i.e.}, $b_s \propto e^{-V_b}$ and  $r_s \propto e^{-V_r}$ [\figref{fig:fig5}(I-b) and \figref{fig:fig5}(I-r)]. However, we can appreciate the distorted/asymmetric shape of the reds' density $r_s$ when size effects are included [\figref{fig:fig5}(III-r)]. Also, in the blues' density we can observe clearly how there is a competition between the potential well and the finite-size repulsion---the particle density $b_s$ inside the well is reduced for finite-size particles. Again, the agreement between the model \eqref{ss_ex1} and the stochastic simulations is excellent. 

\begin{figure*}[htb!]
\begin{center}
\includegraphics[height= 0.43\linewidth]{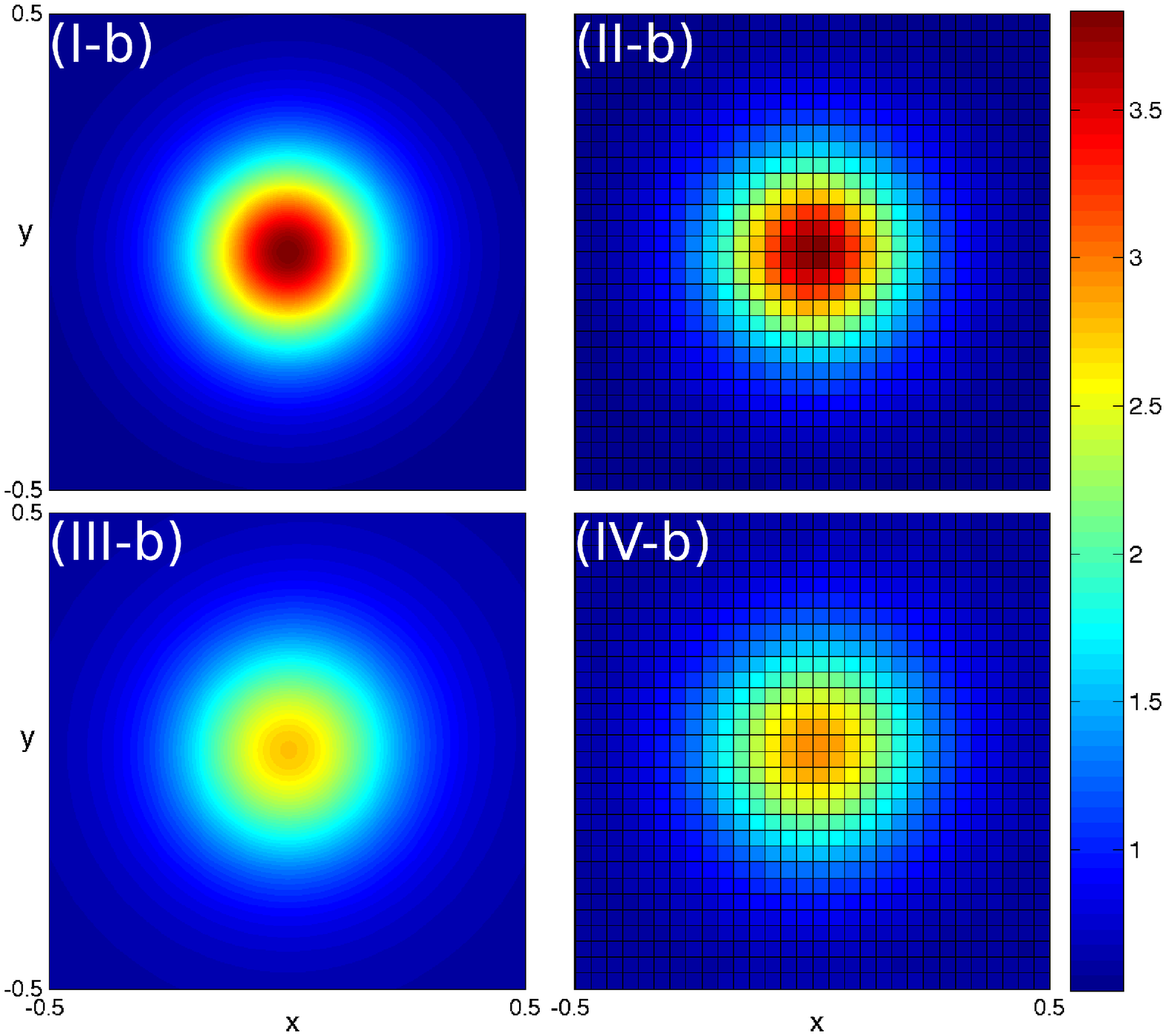}  \includegraphics[height = 0.43\linewidth]{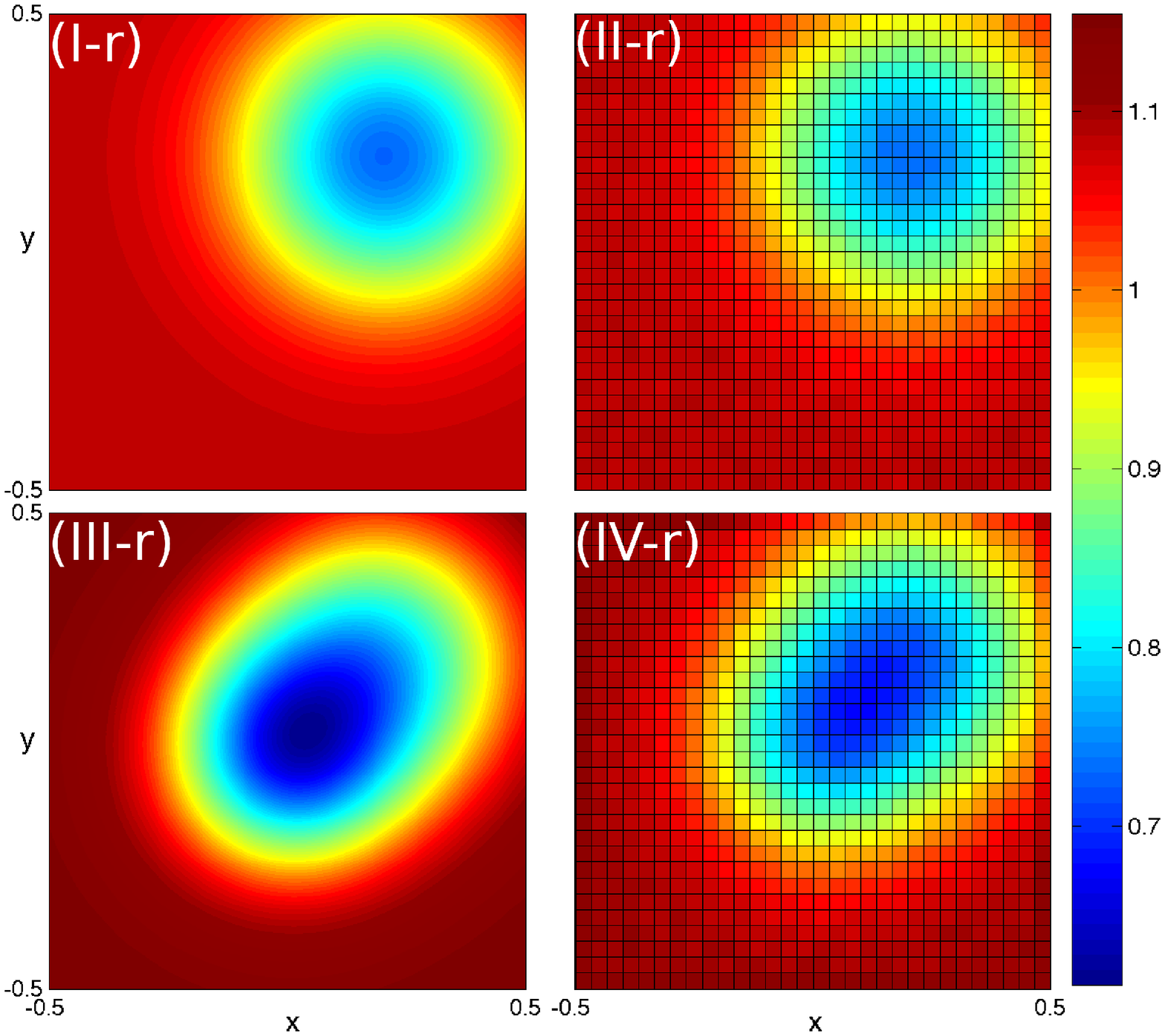}
\caption{Stationary marginal densities $b_s({\bf x})$ ([I--IV]-b) and $r_s({\bf x})$ ([I--IV]-r) for point and finite-size particles, with $V_b = -0.1 \mathcal G({\bf x};0, 0.05)$, $V_r = 0.5 \mathcal G({\bf x};0.2, 0.05)$, $D_b = D_r = 1$ and $N_b=N_r = 400$. 
(I) Solutions of \eqref{ss_ex1} for point particles ($\epsilon_b = \epsilon_r = 0$), $b_s \propto e^{-V_b}$ and  $r_s \propto e^{-V_r}$. (II) Histograms for $\epsilon_b = \epsilon_r = 0$. (III) Solutions of \eqref{ss_ex1} $b_s({\bf x})$ and  $r_s({\bf x})$ for finite-size particles ($\epsilon_b = \epsilon_r = 0.02$). (IV) Histograms for $\epsilon_b = \epsilon_r = 0.02$. Histograms computed from $10^7$ steps of the MH algorithm. All four plots on the left and right have respectively the same color bar.}
\label{fig:fig5}
\end{center}
\end{figure*}

\subsection{Linear stability} \label{sec:linearstab}
It is of interest to compare the upper bounds obtained from the gradient flow structure in \secref{sec:grad flow-entropy} (looking when the mobility matrix $\bf M$ becomes negative definite) with a classical linear stability analysis. We consider a simple case here but the analysis is straightforward for more general cases.

Consider the system \eqref{finalmodel_compactV} with linear potential forces, that is, potential forces of the form $V_b ({\bf x}) = {\bf v}_b \cdot {\bf x}$ and $V_r ({\bf x}) = {\bf v}_r \cdot {\bf x}$. In such cases the equilibrium states are simply $b_s = r_s = 1$. 
We make the following linearization around the equilibrium states, 
\[
b = 1 + \delta A_b \exp \left(\sigma t + i {\bf k \cdot x} \right), \hspace{1.5cm} r = 1 + \delta A_r \exp \left(\sigma t + i {\bf k \cdot x} \right).
\]
Inserting these into \eqref{finalmodel_compactV} and neglecting $\mathcal O(\delta^2)$ terms yields a system ${\bf B}(\sigma, {\bf k}) \left(\begin{smallmatrix}  A_b \\ A_r \end{smallmatrix} \right) = {\bf 0}$. The condition for a non-zero solution, $\det {\bf B} = 0$, is the dispersion relation. For the basic case $\epsilon_b = \epsilon_r = \epsilon$, $D_b = D_r = 1$ and ${\bf v}_b  = {\bf v}_r = 0$, we find that one solution of $\det {\bf B} = 0$ is always negative and the other one is
\begin{equation}
\label{sigmastab}
\sigma({\bf k}) =  \| {\bf k}\|^2   \left\{ -1 + \frac{\epsilon^d \pi}{d} \big[ 2(d-1) + N_b + N_r \big] \right\}.
\end{equation}
This corresponds to a zero-wavelength infinite growth rate instability when
$ \phi + \epsilon^d \pi(d-1)/d>1/2$, where $\phi$ is the particle volume fraction. 

The condition that $\sigma<0$ in \eqref{sigmastab} (for stability) is equivalent to the condition \eqref{upperbound_ex1} in \secref{sec:largenum} found from the mobility matrix under the assumption $N_b, N_r \gg 1$. Observe that in the condition \eqref{upperbound_ex1} the magnitude of the drift terms did not play a role. The numerical exploration of $\text{sgn}(\sigma)$ for several drifts confirms this: an arbitrarily large drift cannot change the sign of $\sigma$. We emphasize again that this instability represents a breakdown of the model reduction, not a true instability in the original particle system. 

\subsection{Symmetrizability of the system and the Onsager relations}
\label{sec:onsager}
In \secref{sec:diffusions} we have seen how our multicomponent diffusion system involves the study of a diffusion matrix that describes how the flux of one component is influenced by its own density gradient and the density gradient of the other component in the system. These ideas can be related with the thermodynamic Onsager reciprocal relations, which establish that in a system fluctuating around its equilibrium a relationship between certain fluxes and thermodynamic forces must hold.\cite{Onsager:1931uu} The Onsager relations are defined assuming the fluctuations around the equilibrium are small (so that the relationship is linear). In particular, if in a system we have the following relations between fluxes $J_i$ and forces $X_i$,
\begin{align}
\label{onsager}
J_i = \sum_k L_{ik} X_k,
\end{align}
the Onsager reciprocal relations requires symmetry in the cross-terms, that is, $L_{ik} = L_{ki}$. The Onsager relations are a macroscopic consequence of  microscopic time reversibility (principle of  detailed balance).\cite{DeGroot:1962ue}$^\text{(p. 35)}$ Note that coefficients $L_{ik}$ can be nonlinear functions of the variables.\cite{DeGroot:1962ue}$^\text{(p. 64)}$ Gupta and Cooper\cite{Gupta:1971vh} study the relationship between the matrix $\bf L$ and the diffusion matrix $\bf D$ in a linear multicomponent diffusion, and find that  $\bf D$ must be positive definite for the Onsager relations to hold. 
It should be pointed out that while the Onsager relations have been named the fourth law in thermodynamics by some authors, their validity has yet to be indisputably established. For instance, many valid multicomponent diffusion models have been found to be inconsistent with these relations.\cite{Zielinski:2001tb}

We proceed to investigate in which cases, if any, our cross-diffusion model \eqref{model_matrix} is consistent with the Onsager relations. It is appropriate to consider the free-energy gradient with respect to the densities to be the force $\bf X$ driving the system to its \emph{minimum} free-energy equilibrium state.\cite{DeGroot:1962ue}$^\text{(p. 35)}$ The gradient-flow structure \eqref{grad_flow} fits with the form required in \eqref{onsager}, and the question is for which cases the mobility matrix ${\bf M}(b,r)$, $\bf L$ in \eqref{onsager}, is symmetric. In \secref{sec:grad flow-entropy} we found two situations for which our system satisfies this: the mobility matrices for the large number of particles' approximation \eqref{mobility_ex1} and for a zero potential \eqref{finalmodel_compact_mobmatrix} (with the appropriate choice of the parameter $\Theta$) are indeed symmetric. 

The fact that in a system with a positive definite diffusion matrix the Onsager relations hold may be related to analytical work on parabolic systems. It is well known that, in hyperbolic or parabolic systems, the existence of a free-energy functional is equivalent to the existence of a change of unknowns which ``symmetrizes'' the system.\cite{Degond:1997uf,Kawashima:1988uk} For parabolic systems, ``symmetrization'' means that the transformed diffusion matrix is symmetric and positive definite.\cite{Hittmeir:2011bd} Note that in \secref{sec:grad flow-entropy} the change of unknowns consisted of going from $(b, r)$ to the so-called \emph{entropy variables} $(\partial_b E, \partial_r E)$.\cite{Burger:2010gb}  The equivalence between Onsager relations and symmetrization comes from the fact that the symmetry is a necessary condition for the entropy production rate to have a sign, itself a condition for the system to be compatible with the second law of thermodynamics.
Although these analytical results seem promising in order to find a free energy for the general form of our system with non-zero potentials \eqref{finalmodel_compactV}, it should be pointed out that finding the change of variables that make our system symmetric (in the sense described in Ref.~\onlinecite{Kawashima:1988uk}) can be in general as challenging as finding the free energy itself.  A first step would be to find a change of unknowns for which the system can be put in a form with no drift terms. 

To conclude this section, we check that the result in Ref.~\onlinecite{Gupta:1971vh} that the original diffusion matrix must be positive definite holds in our case. To $\mathcal O\big(\epsilon_{br}^{d})$, our diffusion matrix \eqref{diffusionmatrix} has eigenvalues 
\begin{equation}
\lambda_i = D_i + \frac{2\pi}{d}  D_i \! \left( (d-1)(N_i-1)\epsilon_i^d \,  i ({\bf x}) - \frac{D_i}{D_i+D_j} N_j \, \epsilon_{br}^d  \, j({\bf x}) \right),
\end{equation}
for $i=b$, $j=r$, and vice versa. A lower bound is $\lambda_i \ge D_i + \tfrac{2\pi}{d} D_i [N_i \epsilon_i^d i - N_j \epsilon_j^d j]$, from which we find that $\lambda_b, \lambda_r > 0$ (since we must have small volume fraction, \emph{i.e.}, $N_b \epsilon_b^d + N_r \epsilon_r^d \ll1$). Therefore, provided there is a small volume fraction, our diffusion matrix is positive definite and hence, as we have already mentioned above, the Onsager relations hold for our system with zero-potentials.  

\section{Conclusions}
\label{sec:conclusions}
In this paper we have considered the diffusion of two interacting species of hard spheres, extending the model derived in Ref.~\onlinecite{Bruna:2012cg} to incorporate particles of different sizes, different diffusivities and under different external forces. The result is a nonlinear cross-diffusion PDE system for the two marginal probability densities associated to each species. 
This approach enables us to describe a complicated system of interacting particles with a simple the continuum PDE model whilst capturing the hard-core interactions at the particle level. These interactions emerge as nontrivial nonlinear terms in the continuum model, involving cross-coupling terms which can be interpreted in terms of the inter-species competition at the population-level. In addition to providing some insight on the system's behavior, the continuum model is relatively easy to solve and analyze. We have assessed the validity of our continuum approach to predict the behavior of the system by comparing its numerical solutions with stochastic simulations of the discrete particle-based model. We have found very good agreement between the two, supporting the idea that by solving a simple system of PDEs we can capture the same population-level behavior observed after many repetitions of expensive stochastic simulations. 

Our analysis is valid in the limit of small but finite particle volume fractions so that pairwise interactions are the dominant ones. This excludes situations close to the jamming limit. Our method uses matched asymptotic expansions in the particle volume fraction to derive the continuum model  in a systematic way as a perturbation of the interaction-free case. Because of the perturbative nature of the method, we expect its accuracy to decline as the volume fraction increases and eventually cease to be valid. In particular, by writing the system in a \emph{gradient-flow} form in terms of the free energy functional and a mobility matrix, we can use the singularity of the mobility matrix as an indicator of the break-down of the model; this occurs at roughly 50\% volume fraction. 
Despite the limitation of a low-volume fraction, we believe our method can provide insight into the mechanisms by which particle-level characteristics emerge at the population-level.

Our method is not based in the thermodynamic limit which requires the number of particles $N$ to tend to infinity together with the system volume.\cite{Felderhof:1978vn} Therefore, the continuum Fokker--Planck (FP) model derived here should not be misinterpreted as a deterministic model for the concentrations of the  two species in the system, valid only when the number of particles $N$ is large. While it can be used in this situation if required, we emphasize that in our work the reduced FP model is valid for any $N$ (one could set $N_b, N_r=1$). In other words, the continuum system is \emph{not} a deterministic model, but rather a PDE system for the \emph{probabilities} of finding a blue and a red particle at a given position and given time.

Our two-component drift--diffusion model captures an enhancement of the collective diffusion rate due to excluded-volume interactions between particles of the same species (as we had already found in our previous work \cite{Bruna:2012cg}), as well as a reduction of the collective diffusion due to interactions with particles of the other species. This structure is useful not only to study the collective diffusion in terms of the particles' volume fraction, but also to analyze the self-diffusion coefficient. The latter describes the evolution of a single tagged particle, and it can be extracted from the model by choosing one of the species to have only one \emph{tagged} individual. In contrast to the collective diffusion, which increases with volume fraction (or relative to point particles), the self-diffusion coefficient decreases with volume fraction. Thus the two species model can be used to characterize transport properties of a system of \emph{identical} particles by distributing them in two subpopulations of $N-1$ and one particles respectively. To our knowledge, such a continuum model capable of explaining both the collective and individual diffusion coefficient has not been reported in the literature/previous works. We have briefly described two experimental procedures to obtain diffusion measurements from real systems, namely a single-particle tracking method to measure the particle's mean-square displacement (MSD) and the ensemble-averaged method FRAP (fluorescence recovery after photobleaching). While it is well understood that the MSD can be related to the self-diffusion coefficient, there is some debate and confusion over the interpretation of FRAP experiments.\cite{Saunders:2012jl} Several methods based on fitting curves to FRAP experimental data exist,\cite{Carrero:2003ea} but none appears to have identified the fact that FRAP is measuring a collective transport property. We believe that our two species model could be used to model the FRAP experiment systematically and provide an improved framework to interpret its results. As can be seen from the FRAP model \eqref{eq_frap},  in general, the measured quantity will not be a pure diffusion coefficient (such as the collective diffusion coefficient, as one might be tempted to think from the fact that FRAP gives an ensemble-averaged measurement) but a mixture of collective diffusion with advection due to gradients in the concentration of the photobleached species. 

We have investigated for which values of the parameters the cross-diffusion system accepts a gradient-flow structure in terms of a free-energy functional; this structure is useful to study the equilibria of the system as well as its stability. Namely, the stationary solutions of the system correspond to the minimizers of the free energy, and the stability can be studied from the convexity of the free energy functional near its equilibria. 

Previous work on the diffusion of two species with size-exclusion interactions using a lattice-based model \cite{Burger:2010gb,Simpson:2009gi} led to a continuum population-level description which is different from our reduced model (which does not restrict the motion of particles to a lattice). In other words, the two approaches (on- and off-lattice models at the particle-level) result in different reduced models, even though they are trying to describe the same problem. It would be interesting to study which rules one needs to prescribe on the lattice model in order to achieve a certain population-level description. We will address this issue in future work. 

The model presented in this work can be extended to consider the diffusion of finite-size particles through obstacles, which has many important applications in porous media and diffusion in biological systems. This may be achieved by setting the diffusivity of one of the species (the obstacles) to zero. An advantage of this approach is that it makes it very easy to study diffusion through spatially varying  concentrations of obstacles.\cite{ValdesParada:2011dr} 
Another interesting extension is to consider anisotropic particles\cite{JabbariFarouji:2012cr} and examine how the continuous PDE model changes with nonspherical particles. 
 
\begin{acknowledgments}
This publication was based on work supported in part by Award No. KUK-C1-013-04, made by King Abdullah University of Science and Technology (KAUST). M.B. acknowledges financial support from EPSRC. The authors also  thank M. Burger for helpful discussions and P. Degond for pointing out the connection between Onsager relations and symmetric systems. 
\end{acknowledgments}


%

\end{document}